\documentclass[journal, twocolumn]{IEEEtran}
\IEEEoverridecommandlockouts

\usepackage{cite}
\usepackage{amsmath,amssymb,amsfonts}
\usepackage{graphicx}
\usepackage{optidef}
\usepackage{textcomp}
\usepackage{xcolor}
\usepackage{multirow}
\usepackage{tikz}
\usetikzlibrary{positioning}
\usepackage{algorithm}
\usepackage{algpseudocode}
\usepackage{float}
\usepackage{booktabs} 
\usepackage{setspace}

\usepackage{siunitx} 
\usepackage{subfig}
\usepackage[draft=true]{hyperref}
\hypersetup{bookmarks=false}
\algrenewcommand\algorithmicindent{1em}
\graphicspath{ {./figures/} }
\def\BibTeX{{\rm B\kern-.05em{\sc i\kern-.025em b}\kern-.08em
    T\kern-.1667em\lower.7ex\hbox{E}\kern-.125emX}}
\usepackage{orcidlink}
\begin{document}

\title{Outage Analysis and Bilevel Resource Optimization in RIS-Assisted Clustered UAV-NOMA Networks}

\author{
    \IEEEauthorblockN{Masoud Ghazikor$^{\orcidlink{0009-0006-2785-362X}}$, \textit{Graduate Student Member, IEEE}, Ly V. Nguyen$^{\orcidlink{0000-0002-2682-4118}}$, \textit{Member, IEEE}, \\ and Morteza Hashemi$^{\orcidlink{0000-0002-9253-6608}}$, \textit{Senior Member, IEEE}}
}

\maketitle
\thispagestyle{plain}
\pagestyle{plain}

\begin{abstract}
This paper investigates reconfigurable intelligent surface (RIS)-assisted downlink uncrewed aerial vehicle (UAV) communications based on non-orthogonal multiple access (NOMA). Since conventional NOMA becomes difficult to scale as the number of UAVs grows, we develop a cluster-based UAV-NOMA framework in which a base station (BS) serves multiple UAV groups with the aid of partitionable RISs. Within this framework, we consider direct, RIS-only indirect, and composite links, and develop an analytical model for the received signal-to-noise ratio (SNR) and outage performance under Nakagami-$m$ and double Nakagami-$m$ fading with imperfect successive interference cancellation (SIC). To enable scalable transmission, we introduce a greedy RIS-aware (GRIS) clustering method, which groups UAVs according to spatial and channel-related features. Based on the resulting clusters and RIS assignment, we formulate an efficiency-reliability bilevel optimization problem that jointly optimizes inter-cluster BS power allocation, intra-cluster NOMA power coefficients, and RIS element assignment. Specifically, the outer efficiency problem maximizes the average throughput across clusters, while the inner reliability problem minimizes the worst-UAV outage probability within each cluster. To solve the formulated problem, we propose a two-level iterative framework consisting of an inter-cluster outer-efficiency solver (IOS) and an intra-cluster inner-reliability solver (I2S), where I2S combines a grid-based progressive search (GPS) for NOMA power refinement with reliability-guided RIS element reassignment, and IOS updates the BS power weights. Numerical results demonstrate the effectiveness of GRIS clustering, while pre-optimization analysis reveals efficiency-reliability trade-offs in power allocation. Baseline comparisons illustrate the ability of IOS-I2S to balance efficiency and reliability, while ablation studies highlight the importance of joint optimization.
\end{abstract}


\renewcommand\thefootnote{}\footnotetext{The material is based upon work supported by NSF grants 1955561, 2212565, 2323189, 2434113, and 2514415. Any opinions, findings, conclusions, or recommendations expressed in this material are those of the author(s) and do not necessarily reflect the views of NSF. \\ The authors are with the Department of Electrical Engineering and Computer Science, The University of Kansas, Lawrence, KS 66045 USA (e-mail: masoudghazikor@ku.edu; vanly.nguyen@ku.edu; mhashemi@ku.edu).}

\section{Introduction}
Uncrewed aerial vehicles (UAVs) are emerging as a key component of future wireless and airspace systems, driven by their expanding roles in public safety, infrastructure inspection, delivery, and commercial applications. This trend is further reflected in current U.S. deployment statistics reported by the Federal Aviation Administration (FAA), which indicate more than 830,000 registered drones and over 480,000 certificated remote pilots as of July 2026 \cite{FAA_UAV_2026}. As UAV density and traffic demand continue to increase, scalable and reliable wireless communication becomes essential for safe and efficient UAV operations, particularly under limited spectrum resources \cite{Chintareddy-2025-Federated}.  

Among the different multiple access technologies considered for dense UAV communications, non-orthogonal multiple access (NOMA) is particularly attractive because it improves spectrum utilization by enabling multiple UAVs to share the same time-frequency resource through power-domain multiplexing \cite{Tan-2019-Downlink}. However, the scalability of UAV-NOMA is fundamentally constrained by successive interference cancellation (SIC), and this challenge becomes even more critical under imperfect SIC conditions, where decoding complexity and error propagation become increasingly significant as the number of served UAVs increases \cite{Lim-2024-Joint}. This limitation is particularly critical in base station (BS)-UAV downlink scenarios, where directly applying NOMA to all UAVs can result in long SIC decoding chains and degraded decoding reliability.

\begin{figure}[t]
\includegraphics[width=0.9\linewidth]{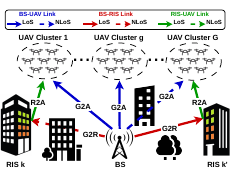}
\centering
\caption{BS serves UAV clusters through three types of links: direct (G2A), RIS-only indirect (G2R2A), and composite (G2A \& G2R2A), while a subset of reflecting elements on passive RIS $k$, denoted by $N_{m_g}^k$ and shown in green, is assigned to UAV $m_g$ in cluster $g$.}
\label{system_model}
\end{figure}

A natural way to improve scalability is to partition UAVs into clusters with shorter NOMA decoding chains. By restricting SIC to a small set of UAVs within each cluster, clustering reduces decoding complexity, mitigates error propagation, and prevents the SIC burden from growing excessively \cite{Liu-2022-Effective}. Moreover, clustering enables a structured transmission framework in which different clusters can be served over orthogonal resource blocks, while NOMA is applied only among UAVs within the same cluster \cite{Zeng-2019-Energy}. As a result, clustering provides an effective mechanism for improving the scalability and reliability of dense UAV-NOMA communications.

Despite the spectral advantages of NOMA, UAV links still suffer from severe path loss, blockage, and highly variable propagation conditions. Reconfigurable intelligent surfaces (RISs) provide a promising solution for improving these links by intelligently shaping the wireless environment through a large number of passive reflecting elements, each of which can adjust its phase shift to enhance constructive signal combining at the receiver \cite{Tabeshnezhad-2026-Jammer}. Since RISs consume only a small amount of power for control and phase-shift tuning, they offer an energy-efficient approach to improving wireless coverage and link reliability \cite{Tang-2025-Throughput}. In clustered UAV-NOMA settings, partitionable RISs are particularly attractive because their reflecting elements can be distributed across clusters and then further assigned to individual UAVs, thereby coupling physical-layer link enhancement with resource allocation decisions \cite{Mohammed-2025-RIS}.

Recent studies have investigated several aspects of RIS-assisted UAV-NOMA communications (see, e.g., \cite{Tang-2025-Throughput, Sobhi-2025-Efficient, Mohammed-2025-RIS, Chen-2025-Secure, Umer-2024-Deep, Goh-2026-Hybrid}). The work in \cite{Tang-2025-Throughput} studies a multi-UAV RIS-assisted NOMA network, where UAVs serve as aerial BSs (ABSs), with user equipment (UE) clustering and joint trajectory, power allocation, and phase shift optimization to maximize throughput, while \cite{Sobhi-2025-Efficient} considers UAVs as either ABSs or RIS carriers and optimizes beamforming, phase shifts, power allocation, and UAV placement for energy efficiency (EE) maximization. More closely related to clustering, \cite{Mohammed-2025-RIS} studies a multi-RIS-assisted UAV-NOMA network with joint UE clustering, link selection, and virtual RIS partitioning to improve UAV connectivity. In addition, \cite{Chen-2025-Secure} investigates a multi-UAV swarm system and \cite{Umer-2024-Deep} considers a NOMA-based UAV-mounted RIS setting, both with joint optimization of UAV trajectories, RIS parameters, and power control, while \cite{Goh-2026-Hybrid} studies a hybrid RIS-UAV downlink NOMA scheme with joint UAV relay placement and power allocation design.

Although RIS-assisted UAV-NOMA networks have been studied, a unified framework for clustered BS-UAV downlink communications with partitionable RISs remains lacking. To address this gap, we develop a RIS-aware UAV clustering algorithm and jointly optimize BS power allocation, NOMA power coefficients, and RIS element assignment under imperfect SIC to balance efficiency and reliability. The main contributions of this paper are summarized as follows:
\begin{itemize}
    \item We present a comprehensive analytical framework to evaluate the outage probability of direct, RIS-only indirect, and composite links in RIS-assisted clustered UAV-NOMA networks under Nakagami-$m$ and double Nakagami-$m$ fading channels. The proposed framework captures the propagation characteristics of the UAV environment, as well as the effects of partitionable RIS assignment and imperfect SIC on outage. In addition, we introduce the greedy RIS-aware (GRIS) clustering based on spatial and channel-related features to enable scalable UAV grouping and intra-cluster NOMA transmission.

    \item Building on the clustered architecture, we formulate an efficiency-reliability bilevel optimization problem in which the outer efficiency level optimizes the inter-cluster BS power allocation, while the inner reliability level optimizes the intra-cluster NOMA power coefficients together with the RIS element assignment. Specifically, the formulation improves efficiency through throughput-oriented cluster-level power allocation while promoting reliability through worst-UAV outage minimization and RIS partitioning within each cluster. To solve this problem, we propose a two-level iterative framework in which the inter-cluster outer-efficiency solver (IOS) handles the outer efficiency problem, the intra-cluster inner-reliability solver (I2S) handles the inner reliability problem, and a grid-based progressive search (GPS) procedure is used within I2S to refine the NOMA power coefficients. 

    \item We evaluate the proposed analytical framework under various system settings and demonstrate that IOS-I2S improves both efficiency and reliability. For example, in the evaluated scenario, IOS-I2S increases average throughput by $12.3\%$ and reduces the network-wide maximum UAV outage by $81.0\%$ compared with equal resource allocation. Ablation results also show that fixing the NOMA power coefficients increases the maximum outage from $0.0909$ to $0.4085$ despite optimizing the other resources, highlighting the importance of joint optimization.
\end{itemize}

This paper extends our preliminary results \cite{Ghazikor-2026-Performance} by developing a comprehensive framework for clustered BS-UAV downlink communications with partitionable RISs under imperfect SIC. We introduce GRIS clustering and inter-cluster BS power allocation and develop IOS-I2S to coordinate resource optimization across clusters. We also expand the numerical evaluation to investigate the effects of clustering weights, RIS element budgets, BS power, and residual SIC, alongside convergence analysis, baseline comparisons, and ablation studies.

The rest of this paper is organized as follows. In Section~\ref{relatedwork}, the related work is presented and discussed. Section~\ref{systemmodel} presents the system model, including the channel models and NOMA throughput analysis. In Section~\ref{clustering}, the proposed UAV clustering method is introduced. Section~\ref{problemformulation} formulates the bilevel efficiency-reliability optimization problem and presents the proposed solution approach. Section~\ref{numericalresults} provides the numerical results, followed by the conclusion in Section~\ref{conclusion}.

\begin{table*}[t]
\scriptsize
\centering
\caption{Summary of related work.}
\label{tab:related_summary}
\renewcommand{\arraystretch}{1.15}
\setlength{\tabcolsep}{3pt}

\begin{tabular}{>{\centering\arraybackslash}p{1.2cm}
                >{\centering\arraybackslash}p{2cm}
                >{\centering\arraybackslash}p{0.5cm}
                >{\centering\arraybackslash}p{0.7cm}
                >{\centering\arraybackslash}p{1cm}
                >{\centering\arraybackslash}p{1.1cm}
                >{\centering\arraybackslash}p{1.6cm}
                >{\centering\arraybackslash}p{1.6cm}
                p{5.3cm}}
\toprule
\textbf{Reference} 
& \textbf{UAV Role(s)} 
& \multicolumn{3}{c}{\textbf{System Components}} 
& \multicolumn{3}{c}{\textbf{Decision Variables}} 
& \centering\textbf{Optimization Objective(s)}\arraybackslash \\
\cmidrule(lr){3-5}
\cmidrule(lr){6-8}
& 
& \textbf{RIS} 
& \textbf{NOMA} 
& \textbf{Clustering} 
& \textbf{BS Power} 
& \textbf{NOMA Power} 
& \textbf{RIS Allocation} 
& \\
\midrule
\cite{Tang-2025-Throughput}
& ABSs
& \checkmark & \checkmark & \checkmark
& $\times$ & \checkmark & $\times$
& Maximize throughput \\
\midrule
\cite{Mohammed-2025-RIS}
& Receivers
& \checkmark & \checkmark & \checkmark
& $\times$ & $\times$ & \checkmark
& Maximize connectivity \\
\midrule
\cite{Sobhi-2025-Efficient}
& ABSs/RIS Carriers
& \checkmark & \checkmark & $\times$
& $\times$ & \checkmark & $\times$
& Maximize EE \\
\midrule
\cite{Chen-2025-Secure}
& Transmitters
& \checkmark & \checkmark & $\times$
& $\times$ & $\times$ & $\times$
& Maximize security rate \\
\midrule
\cite{Umer-2024-Deep}
& RIS Carrier
& \checkmark & \checkmark & $\times$
& $\times$ & \checkmark & $\times$
& Maximize cumulative sum rate \\
\midrule
\cite{Goh-2026-Hybrid}
& Relay/RIS Carrier
& \checkmark & \checkmark & $\times$
& $\times$ & \checkmark & $\times$
& Maximize sum rate \\
\midrule
\cite{Ghazikor-2026-Performance}
& Receivers
& \checkmark & \checkmark & $\times$
& $\times$ & \checkmark & \checkmark
& Minimize maximum outage \\
\midrule
\cite{Lin-2025-Energy}
& ABSs
& \checkmark & $\times$ & \checkmark
& $\times$ & $\times$ & $\times$
& Maximize EE \\
\midrule
\cite{Bhardwaj-2025-Energy}
& RIS Carriers
& \checkmark & $\times$ & \checkmark
& $\times$ & $\times$ & $\times$
& Maximize localization accuracy and SINR \\
\midrule
\cite{Eslami-2026-Hierarchical}
& Comm./Sensing
& \checkmark & \checkmark & \checkmark
& $\times$ & \checkmark & $\times$
& Maximize communication and sensing rates \\
\midrule
\cite{Saif-2025-RIS}
& Receivers
& \checkmark & $\times$ & \checkmark
& $\times$ & $\times$ & \checkmark
& Maximize network connectivity \\
\midrule
\cite{Chen-2024-CoMP}
& Transmitters
& \checkmark & $\times$ & $\times$
& $\times$ & $\times$ & $\times$
& Maximize the sum of the minimum group rates \\
\midrule
\cite{Huroon-2026-Energy}
& Transmitters
& \checkmark & $\times$ & $\times$
& $\times$ & $\times$ & \checkmark
& Maximize EE \\
\midrule
\cite{Katwe-2022-Dynamic}
& ABSs
& $\times$ & \checkmark & \checkmark
& $\times$ & \checkmark & $\times$
& Maximize sum rate throughput \\
\midrule
\cite{Katwe-2022-Energy}
& ABS
& $\times$ & \checkmark & \checkmark
& $\times$ & \checkmark & $\times$
& Maximize EE \\
\midrule
\cite{Yin-2026-Integrated}
& Relay
& $\times$ & \checkmark & \checkmark
& $\times$ & \checkmark & $\times$
& Minimize user transmission latency \\
\midrule
\cite{Zhong-2022-Multi}
& ABSs
& $\times$ & \checkmark & \checkmark
& $\times$ & \checkmark & $\times$
& Maximize throughput \\
\midrule
\cite{Huang-2022-Resource}
& ABS
& $\times$ & \checkmark & \checkmark
& $\times$ & \checkmark & $\times$
& Maximize sum throughput \\
\midrule
\cite{Na-2021-UAV}
& ABS
& $\times$ & \checkmark & \checkmark
& $\times$ & $\times$ & $\times$
& Maximize average sum rate \\
\midrule
\textbf{This work}
& Receivers
& \checkmark & \checkmark & \checkmark
& \checkmark & \checkmark & \checkmark
& Maximize average throughput with outage balancing \\
\bottomrule
\end{tabular}
\end{table*}

\section{Related Work} \label{relatedwork}
In this section, we review the literature on RIS-assisted UAV communications and UAV-NOMA communications. A summary of the related work is provided in Table~\ref{tab:related_summary}.

\subsection{RIS-Assisted UAV Communications}
Recent work on RIS-assisted UAV communications has explored a wide range of system models, UAV roles, optimization objectives, and clustering methods. For example, \cite{Lin-2025-Energy} studies a multi-UAV coverage system, where UAVs serve as ABSs and RIS enhances blocked links. Their framework uses K-density-based clustering for user partitioning and optimizes UAV 3D trajectories, terminal scheduling order, and phase shifts to maximize EE. Also, \cite{Bhardwaj-2025-Energy} investigates a UAV-RIS-assisted IoT network, where UAV-mounted RISs reposition over IoT clusters. Their framework combines K-means clustering with reinforcement learning (RL)-based UAV-RIS relocation to improve localization accuracy and signal-to-interference-plus-noise ratio (SINR) at the BS. Moreover, \cite{Eslami-2026-Hierarchical} studies a multi-UAV RIS-assisted integrated sensing and communication (ISAC) network with uplink NOMA, where UAVs serve users and sense targets. Their framework combines adaptive affinity propagation clustering with optimization of user transmit power, UAV beamforming, and phase shifts to improve communication and sensing rates. Furthermore, \cite{Saif-2025-RIS} considers uplink multi-RIS-assisted UAV communications with UAVs as receivers, where clustering, RIS placement, virtual partitioning, and UE-RIS-UAV link selection are designed to maximize connectivity.

In addition to these clustering-based approaches, other studies consider communication groups without developing a dedicated clustering algorithm. For example, \cite{Chen-2024-CoMP} considers a RIS-assisted multi-UAV multicast system with coordinated UAV transmitters and multicast user groups. Their framework jointly optimizes UAV trajectories, cooperative beamforming, and passive RIS beamforming to maximize the sum of the minimum group rates. Moreover, \cite{Huroon-2026-Energy} studies a beyond-diagonal (BD)-RIS-assisted multi-UAV system, where each transmitter UAV serves a group of users. Their framework optimizes UAV trajectories, BD-RIS element allocation, phase rotations, rate allocation, and precoder design to maximize EE.

\subsection{UAV-NOMA Communications} Similar to RIS-assisted UAV communications, the literature on UAV-NOMA communications spans a broad range of UAV roles, optimization goals, and clustering algorithms. For instance, \cite{Katwe-2022-Dynamic} investigates a multi-UAV-assisted full-duplex (FD)-NOMA cellular network, where UAVs serve as ABSs for uplink/downlink transmissions. Their framework combines K-means clustering with NOMA sub-clustering, and optimizes UAV placement and uplink/downlink power allocation to maximize sum-rate throughput. Similarly, \cite{Katwe-2022-Energy} studies a UAV-assisted FD-NOMA system with adaptive geometric distribution-based user clustering, and optimizes power allocation and UAV placement to maximize EE. In \cite{Yin-2026-Integrated}, a multilayer UAV-enabled NOMA wireless caching network is considered, where a primary UAV coordinates multiple secondary UAVs and may relay content to them. Their framework combines K-means++ clustering with user pairing, power allocation, and UAV deployment to reduce user latency. Moreover, \cite{Zhong-2022-Multi} studies a NOMA-aided multi-UAV cellular network, where UAVs serve as ABSs for offloaded users. Their framework combines K-means clustering with UAV 3D trajectory and power optimization to maximize throughput. In \cite{Huang-2022-Resource}, a UAV serves as an ABS in a multi-cluster NOMA network. Their framework combines K-means clustering and genetic-algorithm-based routing with the optimization of transmission power, hovering locations, and transmission duration to maximize sum throughput. Finally, \cite{Na-2021-UAV} proposes a UAV-supported clustered NOMA system for IoT, where the UAV serves as an ABS. Their framework optimizes UAV trajectory and subslot allocation to maximize the uplink average achievable sum rate.

Building on these studies, our work advances these prior studies by developing a unified system model and analysis framework for UAV clustering, downlink NOMA resource allocation, and partitionable RISs.

\section{System Model} \label{systemmodel}
We consider a single-cell downlink UAV communication system in which a single-antenna BS, assisted by $K$ RISs, serves $M$ single-antenna UAVs using NOMA, where the UAVs and RISs are indexed by $m \in \boldsymbol{M}=\{1,\ldots,M \}$ and $k \in \boldsymbol{K}=\{1,\ldots,K \}$, respectively. To support scalable transmission, the $M$ UAVs are partitioned into $G$ non-overlapping clusters indexed by $g\in\boldsymbol{G}=\{1,\ldots,G\}$, where cluster $g$ contains $M_g$ UAVs with $\sum_{g=1}^{G} M_g = M$. As shown in Fig.~\ref{system_model}, the BS can communicate with each UAV through three links: 1) a direct ground-to-air (G2A) link, 2) an RIS-only indirect ground-to-RIS-to-air (G2R2A) link, and 3) a composite G2A-and-G2R2A link, where each link may experience either line-of-sight (LoS) or non-LoS (NLoS) propagation. Different clusters are served over orthogonal resource blocks, thereby eliminating inter-cluster interference, while UAVs within the same cluster are multiplexed through intra-cluster NOMA using power coefficients collected in $\boldsymbol{\beta}_g=\{\beta_{1},\dots,\beta_{M_g}\}$, with $\sum_{i=1}^{M_g}\beta_i=1$ \cite{Mahmoudi-2022-User, Zeng-2017-Capacity}. For the RIS-assisted links, each cluster is assigned one partitionable RIS, whose reflecting elements are distributed among the UAVs within that cluster; the number of elements assigned from RIS $k$ to UAV $m_g$ is denoted by $N_{m_g}^k$. Accordingly, the BS transmit power is divided across clusters such that cluster $g$ receives a fraction $\omega_g$ of the total power $P^t$, i.e., $P^t_g=\omega_g P^t$, with $\sum_{g=1}^{G}\omega_g=1$. The allocated cluster power is then further distributed among the UAVs in cluster $g$ according to the NOMA power coefficient vector $\boldsymbol{\beta}_g$. In Subsection~\ref{systemmodel}-A, we first establish the channel models and signal-to-noise ratio (SNR) distributions for the RIS-only, direct, and composite links. Subsection~\ref{systemmodel}-B then uses these results to characterize UAV outage probabilities and cluster throughput under imperfect SIC.

\begin{figure*}[t]
\footnotesize
\begin{equation} \label{compositecdf}
F_{\gamma^{c}_{m_g}}(\gamma) \approx
\begin{cases}
\displaystyle
\Big(1 - Q(-\sqrt{a})^{-1} \Big) F_{\gamma^d_{m_g}}(\gamma)
+ \frac{Q(-\sqrt{a})^{-1}}{\Gamma(\hat{m}_3)} \left( \frac{\hat{m}_3}{\Omega_3 |\hat{h}^d_{m_g}|^2} \right)^{\hat{m}_3}
\psi\left(0, \sqrt{\frac{\gamma}{\bar{\gamma}^c_{m_g}}}, \boldsymbol{c} \right),
& \mathbb{E} [\tilde{h}^{r}_{m_g}] > \sqrt{\frac{\gamma}{\Bar{\gamma}^r_{m_g}}} \\[12pt]

\begin{aligned}
\displaystyle
&
\Big(1 - Q(-\sqrt{a})^{-1} \Big) F_{\gamma^d_{m_g}}(\gamma)
+ \frac{Q(-\sqrt{a})^{-1}}{\Gamma(\hat{m}_3)} \, \gamma_{inc}\left(\hat{m}_3, \frac{\hat{m}_3 |\hat{h}^r_{m_g}|^2 (\sqrt{\frac{\gamma}{\bar{\gamma}^r_{m_g}}} - \mathbb{E} [\tilde{h}^{r}_{m_g}])^2}{\Omega_3 |\hat{h}^d_{m_g}|^2} \right) + \frac{Q(-\sqrt{a})^{-1}}{\Gamma(\hat{m}_3)} \times \\ &  \left( \frac{\hat{m}_3}{\Omega_3 |\hat{h}^d_{m_g}|^2} \right)^{\hat{m}_3} \left[ \psi\left( |\hat{h}^r_{m_g}| \Big(\sqrt{\frac{\gamma}{\bar{\gamma}^r_{m_g}}} - \mathbb{E} [\tilde{h}^{r}_{m_g}] \Big), \sqrt{\frac{\gamma}{\bar{\gamma}^c_{m_g}}}, \boldsymbol{c} \right)
- \psi\left(0, |\hat{h}^r_{m_g}| \Big( \sqrt{\frac{\gamma}{\bar{\gamma}^r_{m_g}}} - \mathbb{E} [\tilde{h}^{r}_{m_g}] \Big), \boldsymbol{c} \right) \right],
\end{aligned}
& \mathbb{E} [\tilde{h}^{r}_{m_g}] \leq \sqrt{\frac{\gamma}{\bar{\gamma}^r_{m_g}}}
\end{cases}
\end{equation}

\noindent\rule{\textwidth}{0.8pt}

\begin{equation}
\psi(p_1,p_2,\boldsymbol{c})
=
\sum_{i=0}^{\lfloor 2\hat{m}_3-1 \rceil}
\frac{\rho_i}{2c_1^{\frac{i+1}{2}}}
\begin{cases}
\displaystyle

\Gamma_{inc}\left(
\frac{i+1}{2},c_1\left(p_1-\frac{c_2}{c_1}\right)^2
\right)
-
\Gamma_{inc}\left(
\frac{i+1}{2},c_1\left(p_2-\frac{c_2}{c_1}\right)^2
\right),
& p_1\geq\frac{c_2}{c_1}
\text{ or } i \text{ odd},
\\[12pt]
\displaystyle

\Gamma_{inc}\left(
\frac{i+1}{2},c_1\left(p_2-\frac{c_2}{c_1}\right)^2
\right)
-
\Gamma_{inc}\left(
\frac{i+1}{2},c_1\left(p_1-\frac{c_2}{c_1}\right)^2
\right),
& p_2\leq\frac{c_2}{c_1}
\text{ and } i \text{ even},
\\[12pt]
\displaystyle

\gamma_{inc}\left(
\frac{i+1}{2},c_1\left(p_1-\frac{c_2}{c_1}\right)^2
\right)
+
\gamma_{inc}\left(
\frac{i+1}{2},c_1\left(p_2-\frac{c_2}{c_1}\right)^2
\right),
& p_1<\frac{c_2}{c_1}<p_2
\text{ and } i \text{ even}.
\end{cases}
\end{equation}

\noindent\rule{\textwidth}{0.8pt}

\begin{equation*}
\footnotesize
    \rho_i = {2\hat{m}_3-1 \choose i} \exp{ \Big(\frac{c_2^2}{c_1} - c_3 \Big)} \Big(\frac{c_2}{c_1} \Big)^{2\hat{m}_3 - i - 1}, \quad
    \boldsymbol{c} = (c_1,c_2,c_3) = \Big(\frac{\hat{m}_3}{\Omega_3 |\hat{h}^d_{m_g}|^2} + \frac{1}{|\hat{h}^r_{m_g}|^2 \sigma^2_{\tilde{h}^{r}_{m_g}}}, \frac{\sqrt{\frac{\gamma}{\bar{\gamma}^r_{m_g}}} - \mathbb{E} [\tilde{h}^{r}_{m_g}]}{|\hat{h}^r_{m_g}| \sigma^2_{\tilde{h}^{r}_{m_g}}}, \frac{(\sqrt{\frac{\gamma}{\bar{\gamma}^r_{m_g}}} - \mathbb{E} [\tilde{h}^{r}_{m_g}])^2}{ \sigma^2_{\tilde{h}^{r}_{m_g}}} \Big),
\end{equation*}
\noindent\rule{\textwidth}{0.8pt}
\end{figure*}

\subsection{Channel Modeling}
\subsubsection{RIS-Only Indirect Link} Consider the received power at a UAV $m_g$ in cluster $g$ over the RIS-only indirect link as $P^r_{m_g} = P^t_g |h^r_{m_g}|^2$, where $P^r_{m_g}$ and $P^t_g$ denote the received power and the BS transmit power allocated to cluster $g$, respectively. The equivalent RIS-only channel gain is given by $h^r_{m_g} = \sum_{i=1}^{N^k_{m_g}} h^{gr}_{i} e^{\mathrm{j}\Phi_i^k} h^{ra}_{i}$, where $\Phi_i^k$ denotes the phase shift introduced by the $i$th reflecting element of RIS $k$ \cite{Yang-2020-Performance}. Here, $h^{ra}_{i} = \tilde{h}_i^{ra} e^{\mathrm{j}\phi_i^k} \hat{h}^{ra}$ and $h^{gr}_{i} = \tilde{h}_i^{gr} e^{\mathrm{j}\varphi_i^k} \hat{h}^{gr}$ are the RIS-to-air (R2A) and ground-to-RIS (G2R) channel gains, respectively. Moreover, $\tilde{h}_i$ represents the small-scale fading coefficient modeled by Nakagami-$m$ fading, while $\hat{h} = \sqrt{d^{-\alpha(d)}}$ denotes the square root of the path loss term, which is identical for the $N^k_{m_g}$ elements \cite{Zhang-2022-Joint}. The path loss exponent is expressed as $\alpha(d) = \alpha_{L}\mathbb{P}_{L}(d) + \alpha_{N}(1 - \mathbb{P}_{L}(d))$, where $\alpha_{L}$ and $\alpha_{N}$ denote the LoS and NLoS path loss exponents, respectively. Also, the LoS probability $\mathbb{P}_{L}(d)$ is defined as \cite{Kim-2019-Impact}:
\begin{align*}
\mathbb{P}_{L}(d) = 
\begin{cases}
    \big(1-\exp{(-\frac{z_{1}^2}{2\zeta^2})}\big)^{d\sqrt{\nu\mu}} & \text{$z_{1} = z_{2}$}\\
    \big(1-\frac{\sqrt{2\pi}\zeta}{d^{V}}|Q(\frac{z_{1}}{\zeta})-Q(\frac{z_{2}}{\zeta})|\big)^{d^{H}\sqrt{\nu\mu}} & \text{$z_{1} \neq z_{2}$},
\end{cases}
\end{align*}
where $z_1$ and $z_2$ denote the altitudes of the transmitter and receiver, respectively; $\zeta$, $\nu$, and $\mu$ are environmental parameters; and $Q(\cdot)$, $d^H$, and $d^V$ denote the $Q$-function and the horizontal and vertical distances, respectively. The optimal RIS phase shift is given by $\Phi_i^k = -(\phi_i^k + \varphi_i^k)$ to maximize the received SNR. Thus, the RIS-only channel gain can be rewritten as:
\begin{equation}
    h^{r}_{m_g} = (\hat{h}^{gr} \hat{h}^{ra}) \Big( \sum_{i = 1}^{N^k_{m_g}} \tilde{h}_i^{r} \Big) = (\hat{h}^{r}_{m_g}) (\tilde{h}^{r}_{m_g}),
\end{equation}
where $\tilde{h}_i^{r} = \tilde{h}_i^{gr} \tilde{h}_i^{ra}$ represents the G2R2A fading coefficient associated with the $i$th element. Therefore, $\tilde{h}_{i}^{r}$ is modeled by double Nakagami-$m$ fading, and the PDF of $\tilde{h}_{i}^{r}$ is given by:
\begin{equation}
f_{\tilde{h}_{i}^{r}}(x) = \frac{4x^{\hat{m}_1+\hat{m}_2-1} K_{\hat{m}_1-\hat{m}_2} \Big(2x \sqrt{\frac{\hat{m}_1\hat{m}_2}{\Omega_1 \Omega_2}} \Big) }{\Gamma(\hat{m}_1) \Gamma(\hat{m}_2) (\frac{\Omega_1 \Omega_2}{\hat{m}_1 \hat{m}_2})^{\frac{\hat{m}_1+\hat{m}_2}{2}}},
\end{equation}
where $\Gamma(\cdot)$, $\Omega$, and $K_v(\cdot)$ denote the Gamma function, average fading power, and $v$th-order modified Bessel function of the second kind, respectively \cite{Tegos-2022-Distribution}. The shape parameter $\hat{m}_j$ of channel $j\in\{1,2,3\}$ is approximated as $\hat{m}_j \approx \frac{(\exp(2.708\mathbb{P}_{L}(d_j)^2) + 1)^2}{2 \exp(2.708\mathbb{P}_{L}(d_j)^2) + 1}$ based on the Rician $K$-factor \cite{Channel_Ghazikor_2024}. The $n$th-order moment of $\tilde{h}_{i}^{r}$ is expressed as:
\begin{equation}
    \mathbb{E} [(\tilde{h}_{i}^{r})^n] = \prod_{j=1}^2 \frac{\Gamma(\hat{m}_j + \frac{n}{2})}{\Gamma(\hat{m}_j)} \Big(\frac{\Omega_j}{\hat{m}_j} \Big)^{\frac{n}{2}}.
\end{equation}

\begin{table}[t]
\small
\centering
\caption{Key symbols.}
\label{tab:key_symbols}
\resizebox{\columnwidth}{!}{%
\begin{tabular}{lc}
\toprule
Definition & Symbol \\
\midrule
Number of UAVs and clusters & $M$, $G$ \\
Cluster size and local UAV index & $M_g$, $m_g$ \\
Number of RISs and RIS index & $K$, $k$ \\
Total BS power and cluster transmit power & $P^t$, $P_g^t$ \\
Inter-cluster power vector and cluster weight & $\boldsymbol{\omega}$, $\omega_g$ \\
NOMA power vector and UAV power coefficient & $\boldsymbol{\beta}_g$, $\beta_{m_g}$ \\
RIS element budget assigned to cluster $g$ & $\mathcal{N}_g^k$ \\
RIS assignment vector and UAV element count & $\boldsymbol{N}_g^k$, $N_{m_g}^k$ \\
Direct, RIS-only, and composite channels & $h_{m_g}^d$, $h_{m_g}^r$, $h_{m_g}^c$ \\
Direct, RIS-only, and composite SNRs & $\gamma_{m_g}^d$, $\gamma_{m_g}^r$, $\gamma_{m_g}^c$ \\
Achievable and target rates of UAV $m_g$ & $R_{m_g}$, $\hat{R}_{m_g}$ \\
Residual SIC factor & $\xi$ \\
Outage probability of UAV $m_g$ & $\mathbb{P}_{m_g}^{\mathrm{out}}$ \\
Cluster throughput and average cluster throughput & $R_g$, $\bar{R}$ \\
Minimum total cluster-to-RIS assignment cost & $C_d(\mathcal{C})$ \\
Network-wide minimum channel gain gap & $\delta_h(\mathcal{C})$ \\
RIS proximity and channel disparity weights & $w_d$, $w_h$ \\
\bottomrule
\end{tabular}%
}
\end{table}

By applying the first term of the Laguerre series, the PDF of $\tilde{h}^{r}_{m_g}$ over $N^k_{m_g}$ elements can be approximated as \cite{Abualhayja-2024-Exploiting, Dorra-2011-SIR}:
\begin{equation}
f_{\tilde{h}^{r}_{m_g}}(x) = \frac{x^{a-1} \exp{(-\frac{x}{b})}}{b^a\Gamma(a)}.
\end{equation}
Here, 
\begin{equation}
    a = \frac{\mathbb{E} [\tilde{h}^{r}_{m_g}]^2}{\sigma^2_{\tilde{h}^{r}_{m_g}}} = \frac{N^k_{m_g} \mathbb{E} [\tilde{h}_{i}^{r}]^2}{\mathbb{E} [(\tilde{h}_{i}^{r})^2] - \mathbb{E} [\tilde{h}_{i}^{r}]^2},
\end{equation}
\begin{equation}
    b = \frac{\sigma^2_{\tilde{h}^{r}_{m_g}}}{\mathbb{E} [\tilde{h}^{r}_{m_g}]} = \frac{\mathbb{E} [(\tilde{h}_{i}^{r})^2] - \mathbb{E} [\tilde{h}_{i}^{r}]^2}{\mathbb{E} [\tilde{h}_{i}^{r}]},
\end{equation}
where $\tilde{h}_{i}^{r}$ is independent across $N^k_{m_g}$ elements. Therefore, the probability density function (PDF) and cumulative distribution function (CDF) of the maximum instantaneous SNR
\begin{equation}
    \gamma^{r}_{m_g} = \frac{P^t_g |h^{r}_{m_g}|^2}{P_n} = \frac{P^t_g |\hat{h}^{r}_{m_g}|^2}{P_n} |\tilde{h}^{r}_{m_g}|^2 = \Bar{\gamma}^{r}_{m_g} |\tilde{h}^{r}_{m_g}|^2,
\end{equation}
can be expressed as \cite{Yang-2022-Performance}:
\begin{equation}
f_{\gamma^{r}_{m_g}}(\gamma) = \frac{(\frac{\gamma}{\Bar{\gamma}^{r}_{m_g}})^{\frac{a}{2}-1} \exp{(-\sqrt{\frac{\gamma}{\Bar{\gamma}^{r}_{m_g} b^2}})}}{2\Bar{\gamma}^{r}_{m_g} b^a\Gamma(a)},
\end{equation}
\begin{equation} \label{riscdf}
F_{\gamma^{r}_{m_g}}(\gamma) = \frac{\gamma_{inc}(a, \sqrt{\frac{\gamma}{\Bar{\gamma}^{r}_{m_g} b^2}})}{\Gamma(a)},
\end{equation}
where $\gamma_{inc}(\cdot, \cdot)$ is the lower incomplete Gamma function and $P_n$ denotes the equivalent noise power.

\subsubsection{Direct Link} Similar to RIS-only channel, the direct G2A channel gain $h^{d}_{m_g}$ can be expressed as $h^{d}_{m_g} = \hat{h}^{d}_{m_g} e^{\mathrm{j}\theta_{m_g}} \tilde{h}^{d}_{m_g}$, where $\tilde{h}^{d}_{m_g}$ denotes the small-scale fading coefficient modeled by Nakagami-$m$ fading, whose PDF is given by \cite{Abualhayja-2024-Exploiting}:
\begin{equation}
f_{\tilde{h}^{d}_{m_g}}(x) = \frac{2\hat{m}_3^{\hat{m}_3} x^{2\hat{m}_3-1} \exp{(-\frac{\hat{m}_3x^2}{\Omega_3})}}{\Omega_3^{\hat{m}_3} \Gamma(\hat{m}_3)}.
\end{equation}
Then, the PDF and CDF of the instantaneous SNR, defined as $\gamma^{d}_{m_g} = \frac{P^t_{g} |\hat{h}^{d}_{m_g}|^2}{P_n}|\tilde{h}^{d}_{m_g}|^2 = \bar{\gamma}^{d}_{m_g} |\tilde{h}^{d}_{m_g}|^2$, are given by \cite{Alqahtani-2021-Performance}:
\begin{equation}
f_{\gamma^{d}_{m_g}}(\gamma) = \frac{\hat{m}_3^{\hat{m}_3} (\frac{\gamma}{\bar{\gamma}^{d}_{m_g}})^{\hat{m}_3-1} \exp{(-\frac{\hat{m}_3\gamma}{\Omega_3 \bar{\gamma}^{d}_{m_g}})}}{\bar{\gamma}^{d}_{m_g} \Omega_3^{\hat{m}_3} \Gamma(\hat{m}_3)},
\end{equation} 
\begin{equation}
F_{\gamma^{d}_{m_g}}(\gamma) = \frac{\gamma_{inc}(\hat{m}_3, \frac{\hat{m}_3\gamma}{\Omega_3 \bar{\gamma}^{d}_{m_g}})}{\Gamma(\hat{m}_3)}.
\end{equation}
It is worth noting that the Rayleigh fading for ground-level communication is recovered as a special case when $\hat{m}_3 = 1$.

\begin{figure*}
    \begin{align} \label{modulation_bound}
    \begin{aligned}
        \mathbb{P}(R_{m_g,j} \ge \hat{R}_j) & = \mathbb{P} \Big(\log_2 \big( 1+\frac{\gamma_{m_g} \beta_j}{\gamma_{m_g} \big(\sum_{i=j+1}^{M_g} \beta_i + \xi \sum_{i=1}^{j-1} \beta_i \big) + 1} \big) \ge \hat{R}_j \Big) 
        = \mathbb{P}(\frac{\gamma_{m_g} \beta_j}{\gamma_{m_g} I_j + 1} \ge 2^{\hat{R}_j} - 1) \\ &
        = \mathbb{P}(\gamma_{m_g} \ge \frac{2^{\hat{R}_j} - 1}{\beta_j - (2^{\hat{R}_j} - 1) I_j})
        = \mathbb{P}(\gamma_{m_g} \ge \frac{\tilde{R}_j}{\beta_j - \tilde{R}_j I_j}) = \mathbb{P}(\gamma_{m_g} \ge \hat{\gamma}_j),
    \end{aligned}
\end{align}
\noindent\rule{\textwidth}{0.8pt}
\end{figure*}

\subsubsection{Composite Link} The composite link combines RIS-only indirect and direct links. With the optimal RIS phase shifts given by $\Phi_i^k = -(\phi_i^k + \varphi_i^k - \theta_{m_g})$, the instantaneous SNR of the composite channel can be written as:
\begin{equation}
    \gamma^c_{m_g} = \frac{P^t_g |h^c_{m_g}|^2}{P_n} = \bar{\gamma}^c_{m_g} (|h^{r}_{m_g}| + |h^{d}_{m_g}|)^2,
\end{equation}
where the composite channel gain $h^c_{m_g}$ is formed by the RIS-only indirect $h^r_{m_g}$ and the direct $h^d_{m_g}$ channel gains. Thus, the CDF of the maximum instantaneous SNR $\gamma^{c}_{m_g}$ is defined as:
\begin{equation}
    F_{\gamma^c_{m_g}}(\gamma) = \int^{\sqrt{\frac{\gamma}{\bar{\gamma}^c_{m_g}}}}_{0} F_{|h^r_{m_g}|} \Big(\sqrt{\frac{\gamma}{\bar{\gamma}^c_{m_g}}} - x \Big) f_{|h^d_{m_g}|}(x) dx.
\end{equation}
To improve analytical tractability, the CDF of ${\gamma^r_{m_g}}$ in \eqref{riscdf} is approximated using the $Q$-function as follows:
\begin{equation}
    F_{\gamma^r_{m_g}}(\gamma) \approx 1 - \frac{Q \Big(\frac{\sqrt{\frac{\gamma}{\bar{\gamma}^r_{m_g}}} - \mathbb{E} [\tilde{h}^{r}_{m_g}]}{ \sqrt{\sigma^2_{\tilde{h}^{r}_{m_g}}}} \Big)}{Q(-\sqrt{a})}.
\end{equation}
This approximation is also justified via the central limit theorem in \cite{Analysis_Ni_2023}. Thus, the CDF of $\gamma^c_{m_g}$ can be rewritten as:
\begin{align}
\begin{aligned}
    & F_{\gamma^c_{m_g}}(\gamma) \approx F_{\gamma^d_{m_g}}(\gamma) 
    \\ &
    - \int_{0}^{\sqrt{\frac{\gamma}{\bar{\gamma}^c_{m_g}}}} \frac{Q \Big(\frac{\sqrt{\frac{\gamma}{\bar{\gamma}^c_{m_g}}} - x - |\hat{h}^r_{m_g}| \mathbb{E} [\tilde{h}^{r}_{m_g}]}{|\hat{h}^r_{m_g}| \sqrt{\sigma^2_{\tilde{h}^{r}_{m_g}}}} \Big)}{Q(-\sqrt{a})} f_{|h^d_{m_g}|}(x) dx.
\end{aligned}
\end{align}
Following \cite{Analysis_Ni_2023}, the integral can be solved by dividing the analysis into two cases: $\mathbb{E} [\tilde{h}^{r}_{m_g}] > \sqrt{\frac{\gamma}{\bar{\gamma}^r_{m_g}}}$ and $\mathbb{E} [\tilde{h}^{r}_{m_g}] \le \sqrt{\frac{\gamma}{\bar{\gamma}^r_{m_g}}}$. Therefore, the CDF of $\gamma^c_{m_g}$ is obtained as in \eqref{compositecdf}.

\begin{figure}
    \centering
    \includegraphics[width=0.9\linewidth]{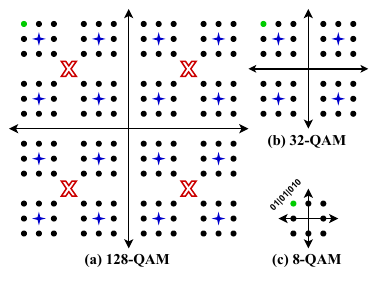}
    \caption{Example of 3 UAVs forming a 128-QAM super-constellation, where the decoding-order constellations are QPSK (UAV 1), QPSK (UAV 2), and 8-QAM (UAV 3). The super-constellation has 128 points with 7 bits per symbol, and the bit format is $b_1b_2|b_3b_4|b_5b_6b_7$.}
    \label{fig:super-constellation}
\end{figure}

\subsection{Effective NOMA Throughput} 
Consider cluster $g$ with $M_g$ UAVs in the downlink NOMA system, where the UAVs are ordered according to their channel gains as $|h_{1}|^2 \le \dots \le |h_{m_g}|^2 \le \dots \le |h_{M_g}|^2$. Accordingly, the NOMA power coefficients are assigned as $\beta_{1} > \dots > \beta_{m_g} > \dots > \beta_{M_g}$, subject to $\sum_{i=1}^{M_g} \beta_i = 1$. Under imperfect SIC, UAV $m_g$ cannot perfectly remove the signals of the first $m_g - 1$ UAVs from the transmitted superimposed signal and therefore treats them as residual interference, while the signals of UAVs with indices $m_g<i\le M_g$ are treated as noise \cite{Chen-2024-Optimizing}. From a modulation perspective, the superimposed signal can also be interpreted through a \emph{super-constellation representation}, as illustrated in Fig.~\ref{fig:super-constellation} \cite{Jafarkhani-2024-Modulation}. The achievable data rate of UAV $m_g$ is given by: 
\begin{equation*}
    R_{m_g} = \log_2 \Big(1+\frac{\gamma_{m_g} \beta_{m_g}}{\gamma_{m_g} \big(\sum_{i=m_g+1}^{M_g}\beta_i + \xi \sum_{i=1}^{m_g-1}\beta_i \big) + 1} \Big),
\end{equation*}
where $\xi \in [0, 1]$ and $\gamma_{m_g} = \frac{\eta_{m_g}}{\beta_{m_g}} = \frac{P_g^t |h_{m_g}|^2}{P_n}$ are the residual SIC factor and the received SNR of the $m_g$-th ordered UAV, respectively. In addition, successful SIC at UAV $m_g$ requires that it can decode the signals intended for all UAVs with indices $j < m_g$ at their target rates before decoding its own signal. Accordingly, $R_{m_g,j}$ represents the data rate at which UAV $m_g$ decodes the signal intended for UAV $j$, where $j \leq m_g$, and is expressed as:
\begin{equation*}
    R_{m_g,j} = \log_2 \Big(1+\frac{\gamma_{m_g} \beta_j}{\gamma_{m_g} \big(\sum_{i=j+1}^{M_g}\beta_i + \xi \sum_{i=1}^{j-1}\beta_i \big) + 1} \Big).
\end{equation*}
By employing order statistics to capture the ordered channel gains in NOMA systems, the PDF and CDF of $\gamma_{m_g}$, for $1 \le m_g \le M_g$, are defined as \cite{Alqahtani-2021-Performance, David-2003-Order}:
\begingroup
\small
\begin{equation*}
    f_{\gamma_{m_g}}(\gamma) = m_g {M_g\choose m_g} f_{\gamma}(\gamma) (F_{\gamma}(\gamma))^{m_g-1} (1-F_{\gamma}(\gamma))^{M_g - m_g},
\end{equation*}
\begin{equation*}
    F_{\gamma_{m_g}}(\gamma) = m_g {M_g\choose m_g} \sum^{M_g - m_g}_{n=0} {M_g - m_g \choose n} \frac{(-1)^n (F_{\gamma}(\gamma))^{m_g+n}}{m_g+n}.
\end{equation*}
\endgroup
Hence, the NOMA-based outage probability of UAV $m_g$, denoted by $\mathbb{P}_{m_g}^{out}(\omega_g, \boldsymbol{\beta}_g, N^k_{m_g})$, is defined as:
\begin{align} \label{noma-gra-out}
\begin{aligned}   
    & \mathbb{P}_{m_g}^{out}(\omega_g, \boldsymbol{\beta}_g, N^k_{m_g}) \\ & = 1 - \mathbb{P} \Big( (R_{m_g,1} \ge \hat{R}_1) \boldsymbol{\cap} \dots \boldsymbol{\cap} (R_{m_g,m_g} \ge \hat{R}_{m_g}) \Big).
\end{aligned} 
\end{align}
Here, $\mathbb{P}(R_{m_g,j} \ge \hat{R}_j)$, for $1 \le j \le m_g$, is simplified as shown in \eqref{modulation_bound}, where $\hat{R}_j$ denotes the target data rate of UAV $j$, $\tilde{R}_j = 2^{\hat{R}_j} - 1$, and $I_j = \sum_{i=j+1}^{M_g} \beta_i + \xi \sum_{i=1}^{j-1} \beta_i$. By defining $\tilde{\gamma}_{m_g} = \max\{\hat{\gamma}_{1}, \dots, \hat{\gamma}_{m_g}\}$ for $m_g<M_g$, and 
$\tilde{\gamma}_{M_g} = \hat{\gamma}_{M_g}$, the outage probability in \eqref{noma-gra-out} is rewritten as:
\begin{align*}
\begin{aligned}
    & \mathbb{P}_{m_g}^{out}(\omega_g, \boldsymbol{\beta}_g, N^k_{m_g}) = 1 - \mathbb{P}(\gamma_{m_g} \ge \tilde{\gamma}_{m_g}) = F_{\gamma_{m_g}}(\tilde{\gamma}_{m_g}) \\ &
    = m_g {M_g \choose m_g} \sum^{M_g - m_g}_{n=0} {M_g - m_g \choose n} \frac{(-1)^n (F_{\gamma}(\tilde{\gamma}_{m_g}))^{m_g+n}}{m_g+n}.
\end{aligned}
\end{align*}

In a delay-limited transmission scheme, the BS transmits at a fixed power $P_t$, and each UAV $m_g$ operates at a predetermined target rate $\hat{R}_{m_g}$. Since the fading channel may not always support these target rates, the effective system throughput depends on the NOMA outage probability $\mathbb{P}_{m_g}^{out}(\omega_g, \boldsymbol{\beta}_g, N^k_{m_g})$ of each UAV $m_g$ \cite{Li-2024-Performance}. Accordingly, the throughput of cluster $g$, denoted by $R_g(\omega_g,\boldsymbol{\beta}_g, \boldsymbol{N}_g^k)$, can be expressed as:
\begin{equation}
    R_g(\omega_g, \boldsymbol{\beta}_g, \boldsymbol{N}_g^k) = \sum_{m_g=1}^{M_g} (1 - \mathbb{P}_{m_g}^{out}(\omega_g, \boldsymbol{\beta}_g, N^k_{m_g})) \ \hat{R}_{m_g},
\end{equation}
where $\boldsymbol{N}_g^k$ denotes the RIS $k$ element assignment vector for $M_g$ UAVs in cluster $g$.

\section{UAV Clustering and RIS Assignment}
\label{clustering}
As the number of UAVs increases, serving all UAVs in a single NOMA group increases decoding complexity and susceptibility to error propagation. To address these problems, the UAVs are partitioned into clusters with separate NOMA decoding chains \cite{Zeng-2017-Capacity, Mahmoudi-2022-User}. In this context, we propose a novel GRIS clustering method that constructs approximately balanced clusters and improves the resulting partition through local refinement, jointly considering UAV-RIS distances and average direct BS-UAV channel gains. The distance criterion favors clusters that can be efficiently assisted by a common RIS, while the channel disparity criterion promotes a larger minimum adjacent channel gain gap across clusters. These larger gaps between average channel gains can reduce the sensitivity of the decoding order to small estimation errors, making it less likely that such errors reverse the relative ordering of UAVs within a cluster. In Subsections~\ref{clustering}-A and~\ref{clustering}-B, we define the RIS assignment cost and channel disparity metric, respectively. Subsection~\ref{clustering}-C combines these metrics into a normalized joint objective, which the GRIS algorithm in Subsection~\ref{clustering}-D seeks to minimize.

\subsection{Cluster-to-RIS Assignment Cost}
Let $\mathcal{C}=\left\{\mathcal{C}_1,\dots,\mathcal{C}_G\right\}$ denote a UAV partition, where $\mathcal{C}_g$ contains the UAVs in cluster $g$. The cost of assigning RIS $k$ to cluster $g$ is the average UAV-RIS distance $c_{g,k}(\mathcal{C})=\frac{1}{|\mathcal{C}_g|}\sum_{m_g\in\mathcal{C}_g}d_{m_g,k}$, where $d_{m_g,k}$ is the corresponding 3D distance. Each cluster is assigned exactly one RIS, and each RIS serves at most one cluster. Let $a_{g,k}\in\{0,1\}$ equal $1$ if RIS $k$ is assigned to serve cluster $g$ and $0$ otherwise. The minimum total assignment cost is then given by:
\begin{align}
C_d(\mathcal{C})
&\triangleq
\min_{\{a_{g,k}\}}
\sum_{g=1}^{G}\sum_{k=1}^{K}
a_{g,k}c_{g,k}(\mathcal{C})
\label{eq:ris_assignment}
\\
\mathrm{s.t.}\quad
&\sum_{k=1}^{K}a_{g,k}=1,\quad
\sum_{g=1}^{G}a_{g,k}\leq1,\quad
a_{g,k}\in\{0,1\},\quad \forall g,k.
\notag
\end{align}
Problem \eqref{eq:ris_assignment} is a linear assignment problem and is solved using the Hungarian algorithm \cite{Kuhn-1955-Hungarian}. Since the cluster composition affects the costs $c_{g,k}(\mathcal{C})$, the RIS assignment is recomputed whenever the candidate UAV partition changes.

\subsection{NOMA Channel Disparity Metric}
For each UAV $m_g\in\mathcal{C}_g$, let
$\bar{h}_{m_g}^{d}=10\log_{10}(\Omega_3|\hat{h}_{m_g}^{d}|^2)$
denote its average large-scale direct channel gain in dB. Within cluster $\mathcal{C}_g$, these gains are ordered from weakest to strongest as $\bar{h}_{g,1}^{d}\leq\dots\leq\bar{h}_{g,|\mathcal{C}_g|}^{d}$, where $\bar{h}_{g,j}^{d}$ denotes the $j$th ordered channel gain in cluster $g$. To avoid grouping UAVs with nearly identical channel conditions, the network-wide channel disparity metric is defined as
the minimum adjacent channel gain gap across all clusters:
\begin{align}
\delta_h(\mathcal{C})
=
\min_{g\in\boldsymbol{G}}
\min_{1\leq j<|\mathcal{C}_g|}
\left(
\bar{h}_{g,j+1}^{d}
-
\bar{h}_{g,j}^{d}
\right).
\label{eq:network_min_gap}
\end{align}
Maximizing $\delta_h(\mathcal{C})$ increases the smallest
adjacent channel gain gap across all clusters, preventing
large gaps in some clusters from masking poor separation
in others.

\subsection{Normalized Joint Objective}
The RIS assignment cost in \eqref{eq:ris_assignment} is measured in meters, whereas the channel disparity metric in \eqref{eq:network_min_gap} is measured in dB. To normalize these metrics, we obtain two reference clusterings for each network realization. The RIS-only reference is constructed to minimize assignment cost; its resulting cost and channel gap are denoted by $C_d^{(D)}$ and $\delta_h^{(D)}$, respectively. Similarly, the channel-only reference is constructed to maximize the minimum channel gap; its resulting cost and channel gap are denoted by $C_d^{(H)}$ and $\delta_h^{(H)}$. These reference values establish the normalization scales. For a candidate clustering $\mathcal{C}$, the normalized RIS assignment penalty is given by:
\begin{align}
\Psi_{d}(\mathcal{C})
=
\frac{
C_{d}(\mathcal{C})-\min\{C_{d}^{(D)},C_{d}^{(H)}\}
}{
\max\{C_{d}^{(D)},C_{d}^{(H)}\}-\min\{C_{d}^{(D)},C_{d}^{(H)}\}
},
\label{eq:normalized_ris_penalty}
\end{align}
and the normalized channel disparity penalty is expressed as:
\begin{align}
\Psi_{h}(\mathcal{C})
=
\frac{
\max\{\delta_{h}^{(D)},\delta_{h}^{(H)}\}
-
\delta_{h}(\mathcal{C})
}{
\max\{\delta_{h}^{(D)},\delta_{h}^{(H)}\}-\min\{\delta_{h}^{(D)},\delta_{h}^{(H)}\}
}.
\label{eq:normalized_diversity_penalty}
\end{align}
Since a lower assignment cost and a larger channel gap are desirable, both \eqref{eq:normalized_ris_penalty} and \eqref{eq:normalized_diversity_penalty} are formulated as minimization penalties. The joint clustering objective is therefore defined as:
\begin{align}
\Psi(\mathcal{C})
=
w_d \Psi_{d}(\mathcal{C})
+
w_h \Psi_{h}(\mathcal{C}),
\label{eq:gris_objective}
\end{align}
where $w_d+w_h=1$ and $w_d,w_h\geq0$. The weights $w_d$
and $w_h$ control the importance of RIS proximity and
channel disparity, respectively. Normalization using
reference values does not guarantee penalties
within $[0,1]$.

\begin{algorithm}[t]
\small
\caption{Greedy RIS-aware (GRIS) Clustering}
\label{alg:gris}
\begin{algorithmic}[1]
\Function{GRIS}{$\boldsymbol{M},G,\tilde{M},
w_d,w_h,
C_d^{(D)},\delta_h^{(D)},
C_d^{(H)},\delta_h^{(H)}$}
    \State $M_{\min}\gets\left\lfloor |\boldsymbol{M}|/G\right\rfloor$, \quad $M_{\max}\gets\left\lceil |\boldsymbol{M}|/G\right\rceil$
    \State $\mathcal{S}\gets\{s_1,\ldots,s_G\}\subseteq\boldsymbol{M}$
    \State $\mathcal{U}\gets\boldsymbol{M}\setminus\mathcal{S}$, \quad $\mathcal{C}_g\gets\{s_g\}$,
    $\forall g\in\boldsymbol{G}$
    \While{$\mathcal{U}\neq\emptyset$}
        \State $n_{\min}
        \gets
        \min_{\{g\in\boldsymbol{G}:|\mathcal{C}_g|<\tilde{M}\}}
        |\mathcal{C}_g|$
        \State $\boldsymbol{G}_{\min}
        \gets
        \{g\in\boldsymbol{G}:
        |\mathcal{C}_g|=n_{\min},
        |\mathcal{C}_g|<\tilde{M}\}$
        \ForAll{$m\in\mathcal{U}$ \textbf{and}
        $g\in\boldsymbol{G}_{\min}$}
            \State $\mathcal{C}^{(m,g)}
            \gets
            \{\mathcal{C}_1,\ldots,
            \mathcal{C}_g\cup\{m\},
            \ldots,\mathcal{C}_G\}$
        \EndFor
        \State $(m^\star,g^\star)
        \gets
        \displaystyle
        \arg\min_{\substack{
        m\in\mathcal{U}, g\in\boldsymbol{G}_{\min}}}
        \Psi(\mathcal{C}^{(m,g)})$
        \State $\mathcal{C}_{g^\star}
        \gets
        \mathcal{C}_{g^\star}\cup\{m^\star\}$, \quad $\mathcal{U}
        \gets
        \mathcal{U}\setminus\{m^\star\}$
    \EndWhile
    \State $\mathrm{improved}\gets\mathrm{true}$
    \While{$\mathrm{improved}$}
        \State $\mathrm{improved}\gets\mathrm{false}$, \quad  $\Delta\Psi_{\max}\gets0$, \quad $\mathrm{type}\gets\mathrm{none}$
        \ForAll{$g_1,g_2\in\boldsymbol{G}$ \textbf{such that} $g_1\neq g_2$}
            \If{$|\mathcal{C}_{g_1}|=M_{\max}$ and
            $|\mathcal{C}_{g_2}|=M_{\min}$}
                \ForAll{$u\in\mathcal{C}_{g_1}$}
                    \State $\mathcal{C}^{(u:g_1\rightarrow g_2)}
                    \gets
                    \{
                    \mathcal{C}_{g_1}\setminus\{u\},
                    \mathcal{C}_{g_2}\cup\{u\}\}$
                    \State $\Delta\Psi
                    \gets
                    \Psi(\mathcal{C})
                    -
                    \Psi(\mathcal{C}^{(u:g_1\rightarrow g_2)})$
                    \If{$\Delta\Psi>\Delta\Psi_{\max}$}
                        \State $\Delta\Psi_{\max}
                        \gets\Delta\Psi$, \quad $\mathrm{type}
                        \gets\mathrm{relocate}$
                        \State $(g_1^\star,g_2^\star,u^\star)
                        \gets(g_1,g_2,u)$
                    \EndIf
                \EndFor
            \EndIf
        \EndFor
        \ForAll{$g_1,g_2\in\boldsymbol{G}$ \textbf{such that} $g_1<g_2$}
            \ForAll{$u\in\mathcal{C}_{g_1}$ and
            $v\in\mathcal{C}_{g_2}$}
                \State $\mathcal{C}^{(u\leftrightarrow v)}
                \gets
                \{
                (\mathcal{C}_{g_1}\setminus\{u\})\cup\{v\},
                (\mathcal{C}_{g_2}\setminus\{v\})\cup\{u\}\}$
                \State $\Delta\Psi
                \gets
                \Psi(\mathcal{C})
                -
                \Psi(\mathcal{C}^{(u\leftrightarrow v)})$
                \If{$\Delta\Psi>\Delta\Psi_{\max}$}
                    \State $\Delta\Psi_{\max}
                    \gets\Delta\Psi$, \quad $\mathrm{type}
                    \gets\mathrm{swap}$
                    \State $(g_1^\star,g_2^\star,
                    u^\star,v^\star)
                    \gets(g_1,g_2,u,v)$
                \EndIf
            \EndFor
        \EndFor
        \If{$\Delta\Psi_{\max}>0$}
            \If{$\mathrm{type}
            =\mathrm{relocate}$}
                \State $\mathcal{C}_{g_1^\star}
                \gets
                \mathcal{C}_{g_1^\star}
                \setminus\{u^\star\}$, \quad $\mathcal{C}_{g_2^\star}
                \gets
                \mathcal{C}_{g_2^\star}
                \cup\{u^\star\}$
            \Else
                \State $\mathcal{C}_{g_1^\star}
                \gets
                (\mathcal{C}_{g_1^\star}\setminus\{u^\star\})\cup\{v^\star\}$
                \State
                $\mathcal{C}_{g_2^\star}
                \gets
                (\mathcal{C}_{g_2^\star}\setminus\{v^\star\})\cup\{u^\star\}$
            \EndIf
            \State $\mathrm{improved}\gets\mathrm{true}$
        \EndIf
    \EndWhile
    \State
    $\{a_{g,k}^{\star}\}
    \gets
    \displaystyle
    \arg\min_{\{a_{g,k}\}}
    \sum_{g=1}^{G}\sum_{k=1}^{K}
    a_{g,k}c_{g,k}(\mathcal{C})$
    s.t.
    \text{Cons.} in \eqref{eq:ris_assignment}
    \State \Return $\mathcal{C}$ and $\{a_{g,k}^{\star}\}$
\EndFunction
\end{algorithmic}
\end{algorithm}

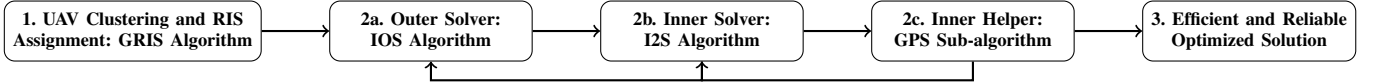
\begin{figure*}[t]
\centering
\resizebox{\textwidth}{!}{
\begin{tikzpicture}[
    node distance=0.9cm and 0.9cm,
    box/.style={
        rectangle,
        rounded corners,
        draw,
        align=center,
        minimum width=2.7cm,
        minimum height=0.8cm,
        font=\scriptsize\bfseries
    },
    arrow/.style={->, thick},
    feedback/.style={->, thick}
]

\node[box] (cluster) {1. UAV Clustering and RIS \\ Assignment: GRIS Algorithm};
\node[box, right=of cluster] (IOS) {2a. Outer Solver: \\ IOS Algorithm};
\node[box, right=of IOS] (i2s) {2b. Inner Solver: \\ I2S Algorithm};
\node[box, right=of i2s] (pgs) {2c. Inner Helper: \\ GPS Sub-algorithm};
\node[box, right=of pgs] (output) {3. Efficient and Reliable \\ Optimized Solution};

\draw[arrow] (cluster) -- (IOS);
\draw[arrow] (IOS) -- (i2s);
\draw[arrow] (i2s) -- (pgs);
\draw[arrow] (pgs) -- (output);

\draw[feedback] (pgs.south) -- ++(0,-0.25) -| (i2s.south);
\draw[feedback] (i2s.south) -- ++(0,-0.25) -| (IOS.south);

\end{tikzpicture}
}
\caption{Overall procedure of the proposed framework.}
\label{fig:overall_flow}
\end{figure*}

\subsection{Greedy RIS-Aware (GRIS) Clustering}
Exhaustively solving the combinatorial clustering problem in \eqref{eq:gris_objective} becomes computationally prohibitive as the number of UAVs increases. To efficiently obtain a high-quality solution, we propose GRIS, summarized in Alg.~\ref{alg:gris}. It seeks to minimize the normalized objective through three stages: greedy cluster construction, local refinement, and RIS assignment.

\subsubsection{Greedy Cluster Construction} Initially, $G$ UAVs are selected as cluster seeds, each initializing one cluster. Let $\mathcal{U}$ denote the set of unassigned UAVs and $\tilde{M}$ the maximum permitted cluster size. At each iteration, $\boldsymbol{G}_{\min}$ denotes the set of clusters having the smallest current cardinality among those satisfying $|\mathcal{C}_g|<\tilde{M}$. These clusters are considered as candidate destinations to maintain balanced cluster sizes. For every unassigned UAV $m\in\mathcal{U}$ and every candidate cluster $g\in\boldsymbol{G}_{\min}$, a candidate clustering $\mathcal{C}^{(m,g)}$ is generated by temporarily assigning UAV $m$ to cluster $g$, and the objective value $\Psi(\mathcal{C}^{(m,g)})$ is evaluated. The candidate assignment yielding the minimum objective is selected as:
\begin{equation}
(m^\star,g^\star)
=
\arg\min_{\substack{
m\in\mathcal{U},\,g\in\boldsymbol{G}_{\min}}}
\Psi(\mathcal{C}^{(m,g)}).
\end{equation}
The selected UAV is assigned to cluster $g^\star$, removed from $\mathcal{U}$, and the procedure repeats until all UAVs have been assigned.

\subsubsection{Local Refinement and RIS Assignment} To further improve the greedy solution, GRIS performs an iterative local refinement stage using two complementary structures. The first structure consists of \emph{balanced relocation} moves, where a UAV is relocated from a cluster of size $M_{\max}$ to a cluster of size $M_{\min}$. Since each relocation moves one UAV between the largest and smallest clusters, the cluster balance is preserved, enabling the algorithm to explore clustering configurations that cannot be reached through pairwise exchanges alone. The second structure consists of \emph{pairwise swaps}, where one UAV from each of two clusters is exchanged while preserving the cardinality of every cluster. For each candidate relocation and swap, the corresponding improvement is computed as $\Delta\Psi=\Psi(\mathcal{C})-\Psi(\mathcal{C}')$, where $\mathcal{C}'$ denotes the candidate clustering after applying the corresponding move. Among all candidate relocations and swaps, the move yielding the largest positive objective improvement is selected and applied. This process repeats until no relocation or swap produces a positive objective improvement, yielding a locally optimal UAV partition. The optimal RIS assignment for this partition is then obtained by solving the binary assignment problem in \eqref{eq:ris_assignment} using the Hungarian algorithm.

\section{Problem Formulation and Proposed Solution}
\label{problemformulation}

\subsection{Bilevel Efficiency-Reliability Problem Formulation} 
We jointly optimize inter-cluster BS power weights, intra-cluster NOMA power coefficients, and RIS element assignment via the following bilevel \emph{efficiency-reliability} problem:
\begin{subequations}
\noindent\rule{0.485\textwidth}{0.8pt}
\begin{align}
\max_{\omega_g} & \ \frac{1}{G} \sum_{g=1}^{G} R_g(\omega_g, \boldsymbol{\beta}_g^{\star}, \boldsymbol{N}_g^{\star}) \label{eq:outer_obj} \\
\text{s.t. }  
& \sum_{g=1}^{G} \omega_g = 1, \quad \omega_g \ge 0, \quad \forall g, \label{eq:outer_con1} \\
& \{\boldsymbol{\beta}_g^{\star}, \boldsymbol{N}_g^{\star}\} \in \arg \min_{\boldsymbol{\beta}_g, \boldsymbol{N}_g^k} \max_{m_g \in \boldsymbol{M}_g} \mathbb{P}^{out}_{m_g} (\omega_g, \boldsymbol{\beta}_g, N_{m_g}^k) \label{eq:inner_obj} \\
& \text{s.t. } \sum_{m_g=1}^{M_g} \beta_{m_g} = 1, \quad \beta_{m_g} \ge 0,\quad \forall m_g, \label{eq:inner_con1} \\ 
& \qquad \tilde{R}_{j} I_{j} < \beta_{j}, \quad 1 \le j \le m_g, \quad \forall m_g, \label{eq:inner_con2} \\
& \quad \ \sum_{m_g=1}^{M_g} N_{m_g}^k \le \mathcal{N}^k_g, \quad N^k_{m_g}\in\mathbb{Z}_{\ge 0}, \quad \forall k \label{eq:inner_con4},
\end{align}
\noindent\rule{0.485\textwidth}{0.8pt}
\end{subequations}
where the outer \emph{efficiency} problem maximizes the average throughput across clusters by optimizing the BS power weights $\{\omega_g\}_{g=1}^G$, subject to \eqref{eq:outer_con1}. For given BS power weights, the inner \emph{reliability} problem determines the optimal intra-cluster variables $\{\boldsymbol{\beta}_g^{\star}, \boldsymbol{N}_g^{\star}\}$ by minimizing the maximum UAV outage probability within each cluster, as stated in \eqref{eq:inner_obj}. Here, $\boldsymbol{\beta}_g$ denotes the NOMA power coefficients for cluster $g$, and \eqref{eq:inner_con1} enforces the NOMA simplex constraint, while \eqref{eq:inner_con2} guarantees imperfect SIC feasibility. Moreover, \eqref{eq:inner_con4} limits the total number of RIS elements assigned to the UAVs in cluster $g$ from RIS $k$, where $\mathcal{N}_g^k$ denotes the preassigned RIS element budget of that cluster. Under these per-cluster RIS budgets, the inner reliability problem becomes separable across clusters, and the BS can solve it in parallel. The resulting $\{\boldsymbol{\beta}_g^{\star}, \boldsymbol{N}_g^{\star}\}$ are then used to evaluate cluster throughput and update the BS power weights.

\subsection{Proposed Bilevel Optimization Framework} \label{proposedsolution}
To solve the formulated bilevel problem, we develop a two-level iterative framework in which the IOS algorithm updates the inter-cluster BS power allocation, while the I2S algorithm jointly optimizes the intra-cluster NOMA power coefficients and RIS element assignment. Fig.~\ref {fig:overall_flow} illustrates the overall procedure of the proposed framework.

\begin{algorithm}[t]
\caption{Grid-based Progressive Search (GPS)}
\small
\label{alg:gps}
\begin{algorithmic}[1]
\Function{GPS}{$\boldsymbol{\beta}_g, \epsilon^{sr}$}
    \If{$\boldsymbol{\beta}_g = \mathrm{none}$}
        \State $\mathcal{C} \gets \left\{ \boldsymbol{\beta}_g \in \boldsymbol{Q}^{M_g} \right\}$, \quad $\mathcal{B}_g \gets \emptyset$
    \Else
        \State $a_{m_g} \gets \max(0, \beta_{m_g} - \epsilon^{sr})$, \quad $b_{m_g} \gets \min(1,  \beta_{m_g} + \epsilon^{sr})$
        \State $\mathcal{C} \gets \prod_{m_g=1}^{M_g} \left( [a_{m_g}, b_{m_g}] \cap \boldsymbol{Q} \right)$, \quad $\mathcal{B}_g \gets \emptyset$
    \EndIf
    \ForAll{$\boldsymbol{\beta}_g \in \mathcal{C}$}
        \If{$|\sum^{M_g}_{m_g=1} \beta_{m_g} - 1| \leq \epsilon$ \textbf{and} $\tilde{R}_{j_g} I_{j_g} < \beta_{j_g}$}
            \State $\mathcal{B}_g \gets \mathcal{B}_g \cup \{ \boldsymbol{\beta}_g \}$
        \EndIf
    \EndFor
    \State \Return $\mathcal{B}_g$
\EndFunction
\end{algorithmic}
\end{algorithm}

\subsubsection{Intra-cluster Inner-reliability Solver (I2S)}
To solve the inner reliability problem for a given BS power weight $\omega_g$, we propose I2S, summarized in Alg.~\ref{alg:i2s}. The algorithm jointly updates the NOMA power coefficient vector $\boldsymbol{\beta}_g$ and the RIS element assignment vector $\boldsymbol{N}_g^k$ for cluster $g$. Initially, the RIS element budget $\mathcal{N}_g^k$ is uniformly assigned among the $M_g$ UAVs, yielding the initial assignment vector $\boldsymbol{N}_g^k$. At each iteration $t$, I2S first updates the NOMA power coefficients using Alg.~\ref{alg:gps} and then refines the RIS element assignment through reliability-guided element reallocation.

\textbf{GPS Sub-algorithm.} GPS generates a candidate set $\mathcal{B}_g$ of feasible NOMA power coefficient vectors with search resolution $\epsilon^{sr}$. When no prior solution is available, GPS performs a global search over the discrete grid $\boldsymbol{Q}^{M_g}$, where $\boldsymbol{Q}=\{0,\epsilon^{sr},2\epsilon^{sr},\dots,1\}$. Otherwise, it constructs a local search space around the current power coefficient vector by restricting each coefficient $\beta_{m_g}$ to the interval $[\max(0,\beta_{m_g}-\epsilon^{sr}),\min(1,\beta_{m_g}+\epsilon^{sr})]$ and intersecting this interval with the global grid. Each candidate vector is then tested against the NOMA simplex and SIC feasibility constraints, and only the feasible candidates are retained in $\mathcal{B}_g$.

\begin{algorithm}[t]
\caption{Intra-cluster Inner-reliability Solver (I2S)}
\label{alg:i2s}
\begin{algorithmic}[1]
\small
\Function{I2S}{$\omega_g$, $\lambda$, $\epsilon^{in}$, $\epsilon^{ac}$}

\State $N_{m_g}^k \gets 
\left\lfloor\frac{\mathcal{N}_g^k}{M_g}\right\rfloor,\quad
\boldsymbol{N}_g^k
\gets [N_{m_g}^k]_{m_g=1}^{M_g},\quad
\boldsymbol{\beta}_g^0 \gets \mathrm{none}$

\For{$t=1$ to $T_i$}

\State $\epsilon^{sr} \gets \epsilon^{in},\quad
\boldsymbol{\beta}_g^t \gets \mathrm{none},\quad
\boldsymbol{N}_g^{\star} \gets \boldsymbol{N}_g^k$

\While{$\epsilon^{sr}>\epsilon^{ac}$}

\State $\mathcal{B}_g
\gets \Call{GPS}{\boldsymbol{\beta}_g^t,\epsilon^{sr}}$

\State $\boldsymbol{\beta}_g^t
\gets
\arg\min_{\boldsymbol{\beta}_g\in\mathcal{B}_g}
\max_{m_g\in\boldsymbol{M}_g}
\mathbb{P}_{m_g}^{out}
(\omega_g,\boldsymbol{\beta}_g,N_{m_g}^k)$

\State $\epsilon^{sr}
\gets \lambda\epsilon^{sr}$

\EndWhile

\While{
$\max\limits_{m_g}
\mathbb{P}_{m_g}^{out}
(\cdot,\boldsymbol{\beta}_g^t,N_{m_g}^{\star})
\le
\max\limits_{m_g}
\mathbb{P}_{m_g}^{out}
(\cdot,\boldsymbol{\beta}_g^t,N_{m_g}^k)$
}

\State $\boldsymbol{N}_g^{k}
\gets \boldsymbol{N}_g^{\star}$, $m_g^{\max}
\gets
\arg\max_{m_g}
\mathbb{P}_{m_g}^{out}
(\omega_g,\boldsymbol{\beta}_g^t,N_{m_g}^{\star})$

\State $\mathcal{U}_g
\gets
\{u_g\in\boldsymbol{M}_g:
u_g\neq m_g^{\max},\,
N_{u_g}^{\star}\ge1\}$

\If{$\mathcal{U}_g\neq\emptyset$}

\ForAll{$u_g\in\mathcal{U}_g$}

\State $\boldsymbol{N}_g^{(u_g)}
\gets \boldsymbol{N}_g^{\star},\quad
\boldsymbol{N}_g^{(u_g)}[u_g]
\gets N_{u_g}^{\star}-1$

\State $\boldsymbol{N}_g^{(u_g)}[m_g^{\max}]
\gets N_{m_g^{\max}}^{\star}+1$

\EndFor

\State $u_g^{\star}
\gets
\arg\min\limits_{u_g\in\mathcal{U}_g}
\max\limits_{m_g\in\boldsymbol{M}_g}
\mathbb{P}_{m_g}^{out}(
\omega_g,\boldsymbol{\beta}_g^t,
\boldsymbol{N}_g^{(u_g)}[m_g])$

\State $N_{u_g^{\star}}^{\star}
\gets N_{u_g^{\star}}^{\star}-1,\quad
N_{m_g^{\max}}^{\star}
\gets N_{m_g^{\max}}^{\star}+1$

\Else

\State \textbf{break}

\EndIf

\EndWhile

\If{
$t > 1$ \textbf{and} $\|\boldsymbol{\beta}_g^t
-\boldsymbol{\beta}_g^{t-1}\|
<\epsilon$  
}

\State \textbf{break}

\EndIf

\State $\boldsymbol{\beta}_g^{t-1}
\gets \boldsymbol{\beta}_g^t$

\EndFor

\State \textbf{return}
$\boldsymbol{\beta}_g^{\star}
\gets\boldsymbol{\beta}_g^t,\quad
\boldsymbol{N}_g^{\star}
\gets \boldsymbol{N}_g^{k}$

\EndFunction
\end{algorithmic}
\end{algorithm}

\textbf{I2S Algorithm.}
Given $\mathcal{B}_g$, I2S selects the NOMA power coefficient vector $\boldsymbol{\beta}_g^t$ that minimizes the maximum UAV outage probability within cluster $g$. The search resolution is then progressively reduced by the refinement factor $\lambda$ until it reaches the accuracy threshold $\epsilon^{ac}$, providing a coarse-to-fine refinement of the power coefficients. After updating $\boldsymbol{\beta}_g^t$, I2S refines the RIS element assignment through a reallocation process. At each step, the UAV with the largest outage probability is identified as the receiver of one additional RIS element. All other UAVs with at least one assigned RIS element are considered as "candidate donors", and a candidate assignment is constructed for each possible one-element transfer. The donor is then selected as the one that minimizes the resulting maximum UAV outage probability. The corresponding transfer is accepted as long as the resulting maximum outage does not exceed that of the previously accepted assignment, and the process is repeated until no further non-worsening reassignment is available. The I2S iterations terminate when the NOMA power coefficient vector converges, i.e., $\|\boldsymbol{\beta}_g^t-\boldsymbol{\beta}_g^{t-1}\|<\epsilon$. At convergence, I2S returns the optimized NOMA power coefficient vector $\boldsymbol{\beta}_g^{\star}$ and RIS element assignment vector $\boldsymbol{N}_g^{\star}$.

\begin{figure*}[t]
    \centering
    \begin{minipage}{.33\linewidth}
    \includegraphics[width=6cm, trim={270 0 270 0}, clip]{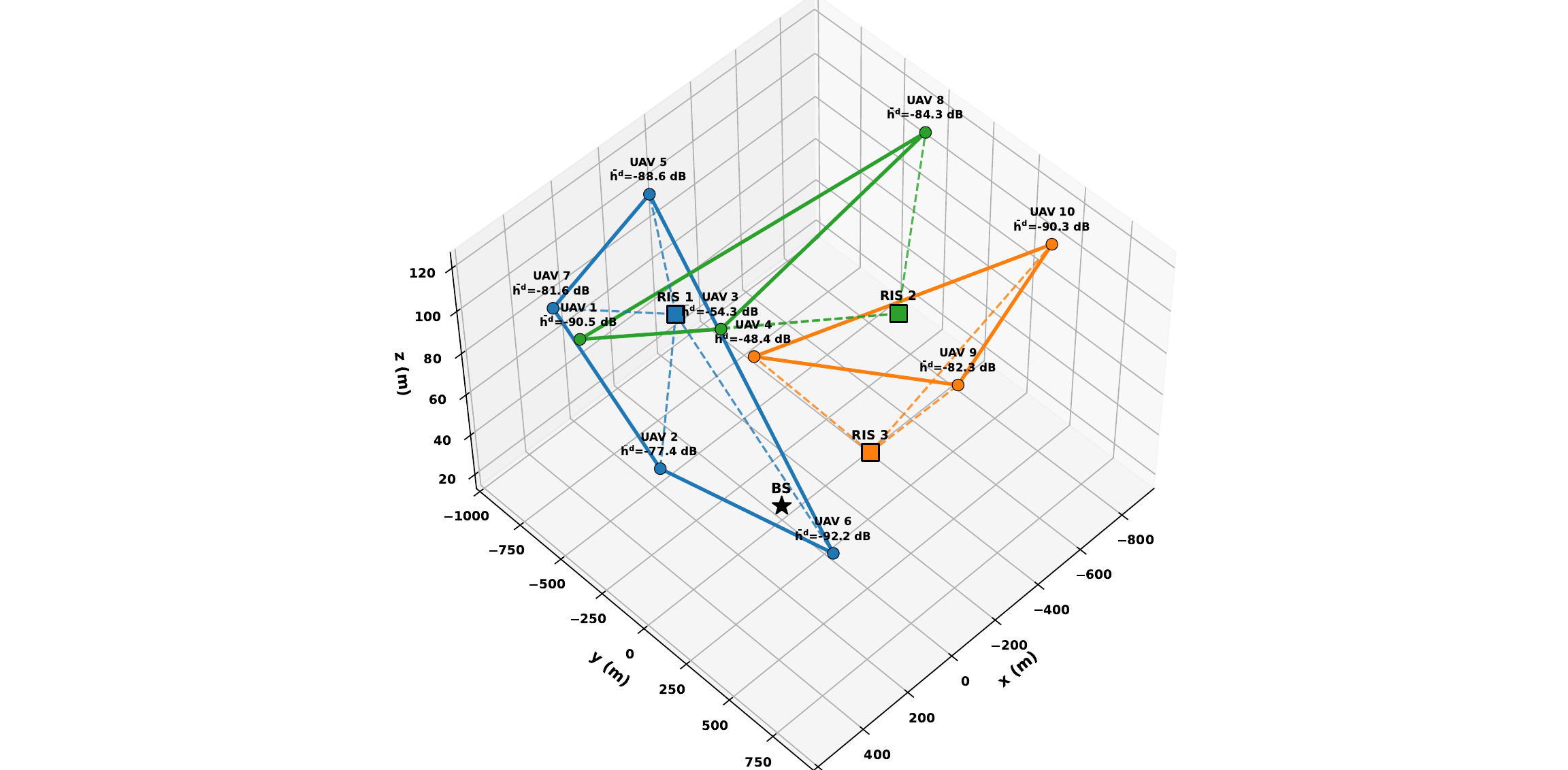}
        \centering (a) $w_d$ = 0.0, $w_h$ = 1.0
    \end{minipage}%
    \begin{minipage}{.33\linewidth}
   \includegraphics[width=6cm, trim={270 0 270 0}, clip]{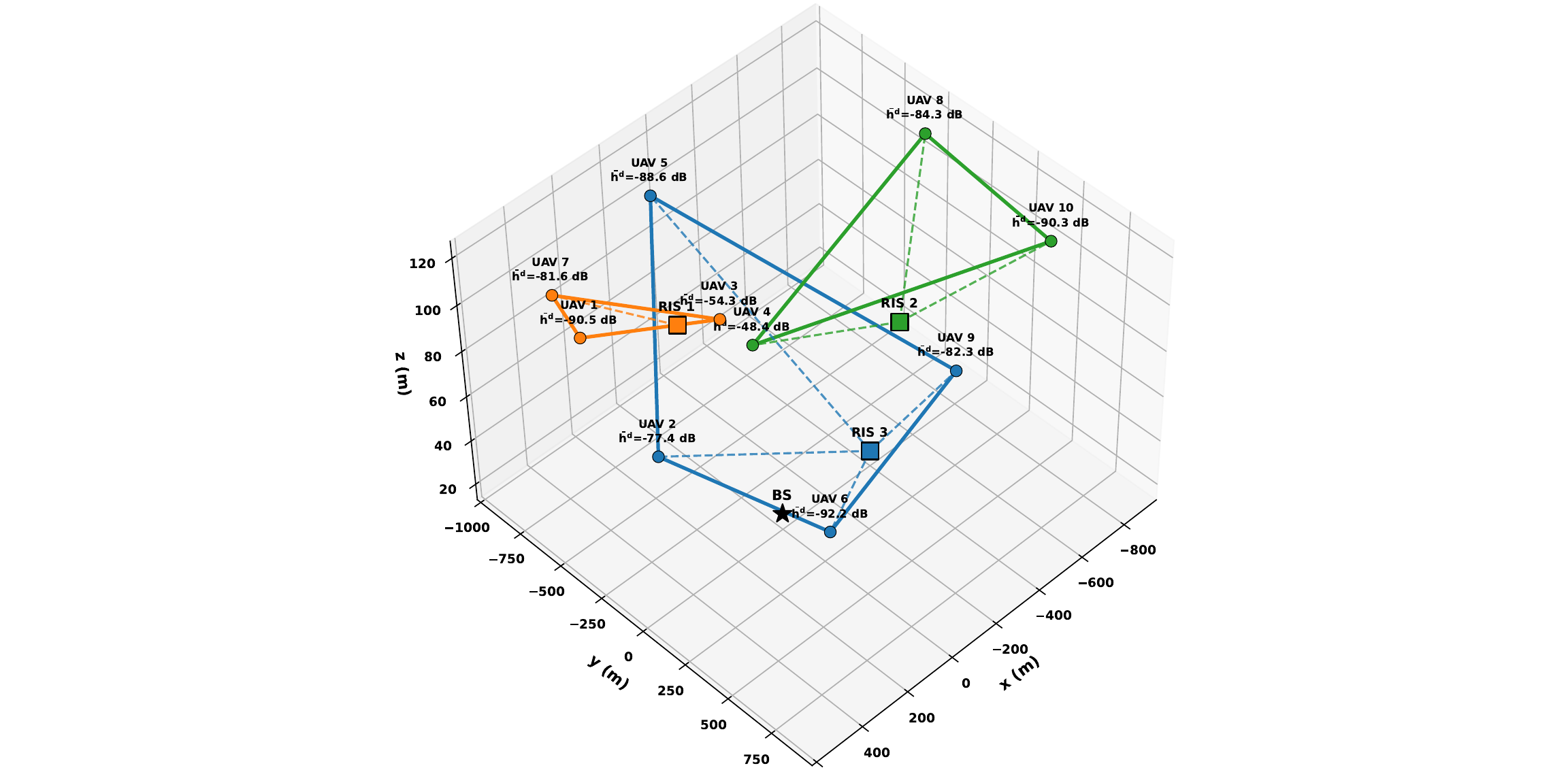}
        \centering (b) $w_d = w_h$ = 0.5
    \end{minipage}%
    \begin{minipage}{.33\linewidth}
    \includegraphics[width=6cm, trim={270 0 270 0}, clip]{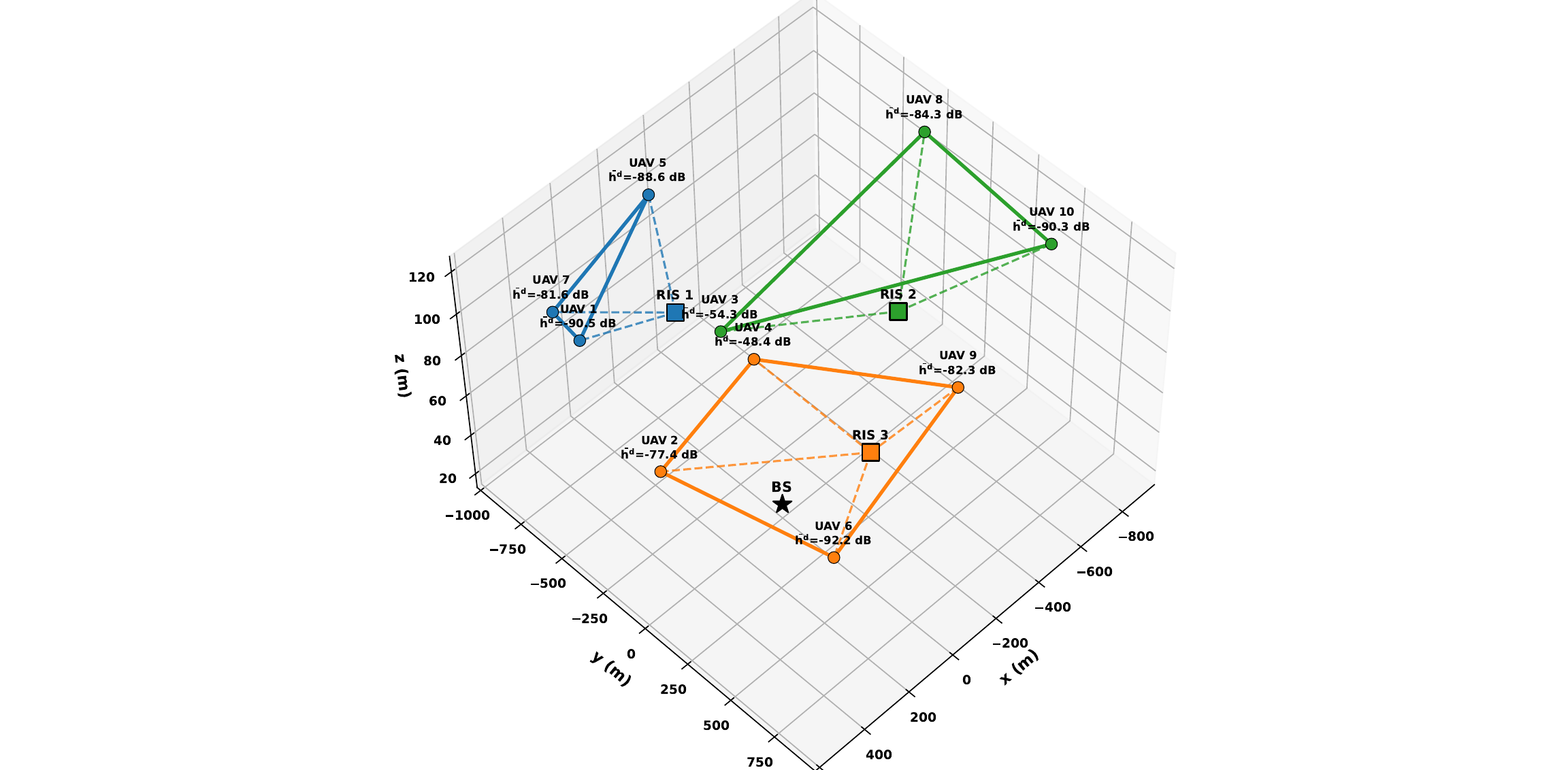}
        \centering (c) $w_d$ = 1.0, $w_h$ = 0.0
    \end{minipage}%
    \caption{Impact of the RIS proximity weight $w_d$ and channel disparity weight $w_h$ on GRIS clustering and RIS assignment.}
\label{Fig:3D_gris_clustering}
\end{figure*}

\begin{algorithm}[t]
\caption{Inter-cluster Outer-efficiency Solver (IOS)}
\small
\label{alg:ios}
\begin{algorithmic}[1]
\Function{IOS}{$G,\alpha,\epsilon^{fd},\Delta, \chi, \lambda, \epsilon^{in}, \epsilon^{ac}$}
    \State $\boldsymbol{\omega} \gets \left[\tfrac{1}{G},\dots,\tfrac{1}{G}\right]$,
    \quad $\{\boldsymbol{\beta}^{\star}_g,\boldsymbol{N}^{\star}_g\}_{g=1}^G \gets \textsc{I2S}(\boldsymbol{\omega}, \lambda, \epsilon^{in}, \epsilon^{ac})$
    \State $\bar{R} \gets \frac{1}{G}\sum_{g=1}^G R_g(\omega_g, \boldsymbol{\beta}^{\star}_g,\boldsymbol{N}^{\star}_g)$
    \For{$t=1$ to $T_o$}
        \For{$i=1$ to $G$}
            \State $\boldsymbol{\omega}^{(i)} \gets \boldsymbol{\omega}, \quad \omega^{(i)}_i \gets \omega_i + \epsilon^{fd}$,
            \quad $\boldsymbol{\omega}^{(i)} \gets \frac{\boldsymbol{\omega}^{(i)}} {\sum_{g=1}^G \omega^{(i)}_g}$
            \State $\{\boldsymbol{\beta}^{(i)}_g,\boldsymbol{N}^{(i)}_g\}_{g=1}^G \gets \textsc{I2S}(\boldsymbol{\omega}^{(i)}, \lambda, \epsilon^{in}, \epsilon^{ac})$
            \State $\bar{R}^{(i)} \gets \frac{1}{G}\sum_{g=1}^G R^{(i)}_g (\omega_g^{(i)}, \boldsymbol{\beta}^{(i)}_g,\boldsymbol{N}^{(i)}_g)$
            \State $\widehat{\nabla}_i \bar{R} \gets (\bar{R}^{(i)}-\bar{R})/\epsilon^{fd}$
        \EndFor
        \For{$g=1$ to $G$}
            \State $\omega_g \gets \omega_g \exp\!\big(\frac{\alpha}{t} \mathrm{clip}(\widehat{\nabla}_g \bar{R},-\chi,\chi)\big)$
        \EndFor
        \State $\boldsymbol{\omega} \gets \frac{\boldsymbol{\omega}} {\sum_{g=1}^G \omega_g}$,
        \quad $\{\boldsymbol{\beta}^{\star}_g,\boldsymbol{N}^{\star}_g\}_{g=1}^G \gets \textsc{I2S}(\boldsymbol{\omega}, \lambda, \epsilon^{in}, \epsilon^{ac})$
        \State $\bar{R}^{new} \gets \frac{1}{G}\sum_{g=1}^G R_g(\omega_g, \boldsymbol{\beta}^{\star}_g,\boldsymbol{N}^{\star}_g)$

        \If{$|\bar{R}^{new}-\bar{R}| < \Delta$}
            \State $\boldsymbol{\omega}^{\star} \gets \boldsymbol{\omega}$
            \State \textbf{break}
        \EndIf
        \State $\bar{R} \gets \bar{R}^{new}$, \quad $\boldsymbol{\omega}^{\star} \gets \boldsymbol{\omega}$
    \EndFor
    \State \Return $\{\omega^{\star}_g,\boldsymbol{\beta}^{\star}_g,\boldsymbol{N}^{\star}_g\}_{g=1}^G$
\EndFunction
\end{algorithmic}
\end{algorithm}

\subsubsection{Inter-cluster Outer-efficiency Solver (IOS)}
To solve the outer efficiency problem over the simplex-constrained BS power weight vector $\boldsymbol{\omega}=[\omega_1,\dots,\omega_G]$, we propose IOS shown in Alg.~\ref{alg:ios}. Starting from a uniform initialization, IOS evaluates the average cluster throughput as:
\begin{align}
\bar{R}(\boldsymbol{\omega})
=
\frac{1}{G}\sum_{g=1}^{G}
R_g(\omega_g,\boldsymbol{\beta}_g^{\star},\boldsymbol{N}_g^{\star}),
\end{align}
where, for a given $\boldsymbol{\omega}$, the inner solver I2S returns the corresponding intra-cluster variables $\{\boldsymbol{\beta}_g^{\star},\boldsymbol{N}_g^{\star}\}$ for each cluster $g$. Since the mapping from $\boldsymbol{\omega}$ to $\bar{R}(\boldsymbol{\omega})$ is available only implicitly through I2S, the outer problem is treated as a black-box optimization problem over the simplex.

To update $\boldsymbol{\omega}$, IOS estimates the sensitivity of $\bar{R}(\boldsymbol{\omega})$ to each BS power weight using finite differences. Specifically, for each coordinate $i$, the algorithm constructs a perturbed feasible vector $\boldsymbol{\omega}^{(i)}$ by increasing $\omega_i$ by $\epsilon^{fd}$ and then renormalizing the resulting vector to satisfy the simplex constraint. The inner solver I2S is subsequently invoked under this perturbed allocation, yielding the corresponding average throughput $\bar{R}^{(i)}$. The finite difference estimate $\widehat{\nabla}_i \bar{R} = (\bar{R}^{(i)}-\bar{R})/\epsilon^{fd}$ measures the local sensitivity of the average throughput to increasing the power weight of cluster $i$. These estimates are then used in an exponentiated gradient update, where each weight $\omega_g$ is multiplied by an exponential factor determined by the corresponding clipped gradient value. The clipping operation, controlled by the threshold $\chi$, prevents excessively large updates and improves numerical stability. After the multiplicative update, the weight vector is renormalized to preserve simplex feasibility. The updated weights are then used to recompute the inner solutions and the corresponding average throughput. This iterative process continues until the change in average throughput falls below the tolerance $\Delta$, i.e., $|\bar{R}^{new}-\bar{R}|<\Delta$.

\begin{table} [t]
    \centering
    \small
    \caption{Key simulation parameters.}
    \resizebox{\columnwidth}{!}{%
    \label{tab:sim_parameters}
    \begin{tabular}{lc}
        \toprule
        Definition & Notation \& Value \\
        \midrule
        Cell Radius & $r = 2$ km \\
        Environmental Parameters & $\zeta = 20$, $\nu = 3 \times 10^{-4}$, $\mu = 0.5$\\
        Path Loss Exponent & $\alpha_{L}=2$, $\alpha_{N}=3.5$\\
        Average Fading Power & $\Omega=1$\\
        Noise Power & $P_n\approx-103.98$ dBm \\
        BS Transmit Power & $P_t = 10$ dBm\\
        Target Data Rate & $\hat{R}_{m_g}= 2$ bpc\\
        Residual SIC Factor & $\xi= 0.001$ \\
        Maximum RIS Elements & $\mathcal{N}^k = 12 \times 12$\\
        I2S Hyperparameters & $\lambda=0.9, \epsilon^{in}=0.1,  \epsilon^{ac}=10^{-8}$ \\
        IOS Hyperparameters & $\alpha=0.5, \epsilon^{fd}=0.05, \Delta = 10^{-3}$\\
        \bottomrule
    \end{tabular}
    }
\end{table}

\subsubsection{Time Complexity of the Proposed Algorithms}
Assuming negligible communication and scheduling overhead, let $W_c\leq G$ clusters run concurrently with $W_p$ workers each, where $W_cW_p\leq W$ for $W$ available workers. Set $L=\max_g M_g$, and let $A,B$ bound total and feasible GPS candidates per level. Let $S,H$ bound GPS levels and RIS-transfer rounds per I2S iteration, and $T_r$ count GRIS refinement iterations. Let $C_o,C_\Psi$ denote complete cluster-outage and clustering-objective evaluation costs. GPS screening and parallel candidate evaluation cost $C_\beta=\mathcal{O}(AL^2+\lceil B/W_p\rceil C_o)$ per level, yielding $C_i=\mathcal{O}(T_i[SC_\beta+H\lceil(L-1)/W_p\rceil(L+C_o)])$ per I2S solve with parallel donor evaluation. The overall complexity combines the IOS cost $\mathcal{O}([1+T_o(G+1)]\lceil G/W_c\rceil C_i+T_oG^2)$ with the one-time sequential GRIS cost $\mathcal{O}(M^2(G+T_r)C_\Psi)$.

\section{Numerical Results} \label{numericalresults}
In our simulations, 10 UAVs and 3 RISs are deployed within the BS-centered cell via Poisson point processes (PPP). Key simulation parameters are summarized in Table \ref{tab:sim_parameters}. We first evaluate GRIS clustering, then examine the pre-optimization performance, and finally present the optimization results.

\begin{figure*}[t]
    \centering
    \includegraphics[width=0.75\linewidth]{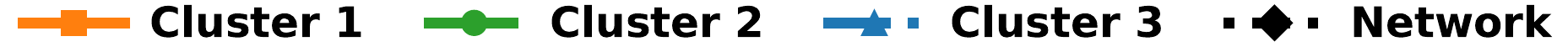}
    \begin{minipage}{.33\linewidth}
    \includegraphics[width=6cm]{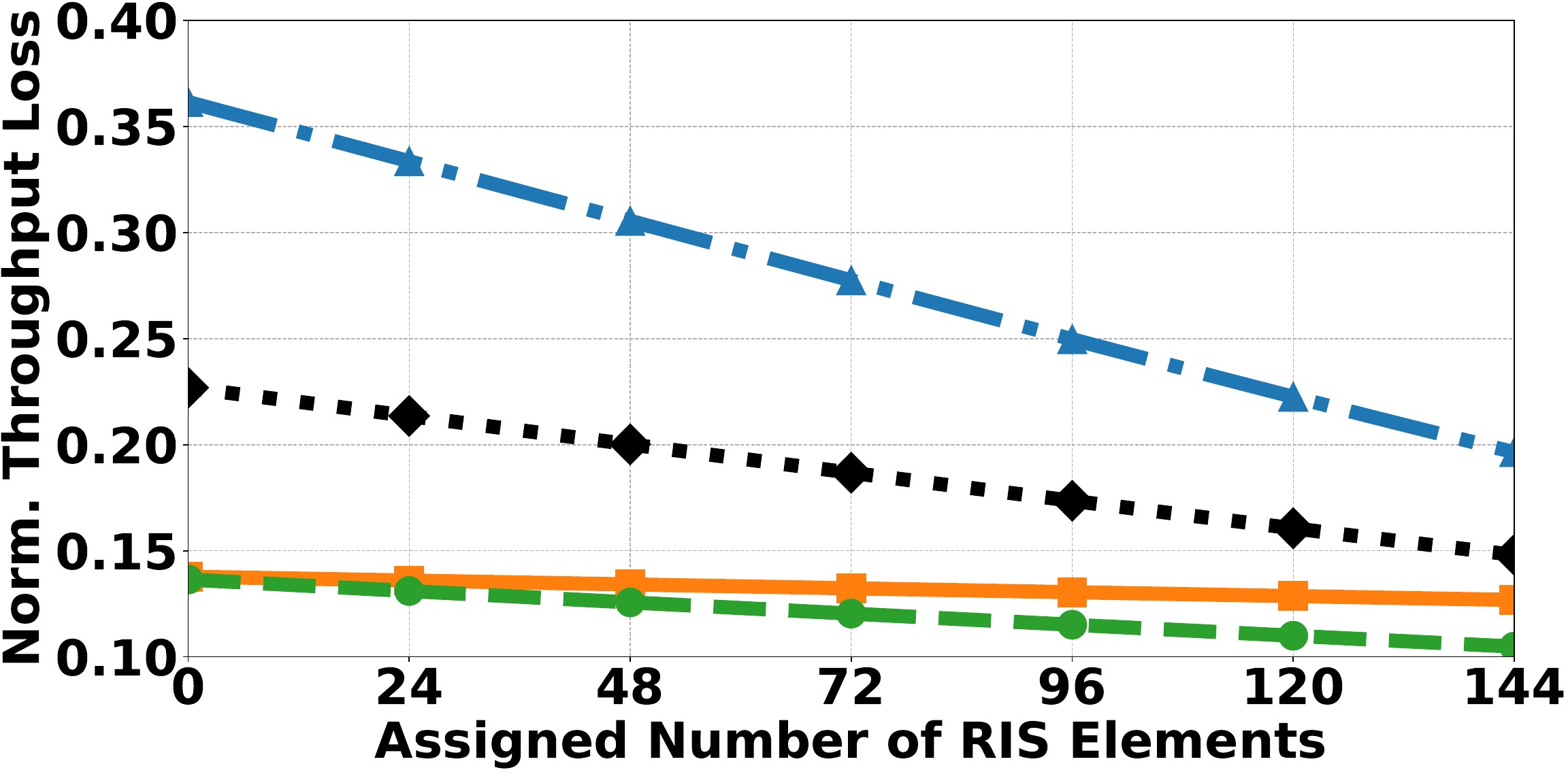}
        \centering (a) $\mathcal{L}_g$ vs. $N^k_g$ 
    \end{minipage}%
    \begin{minipage}{.33\linewidth}
   \includegraphics[width=6cm]{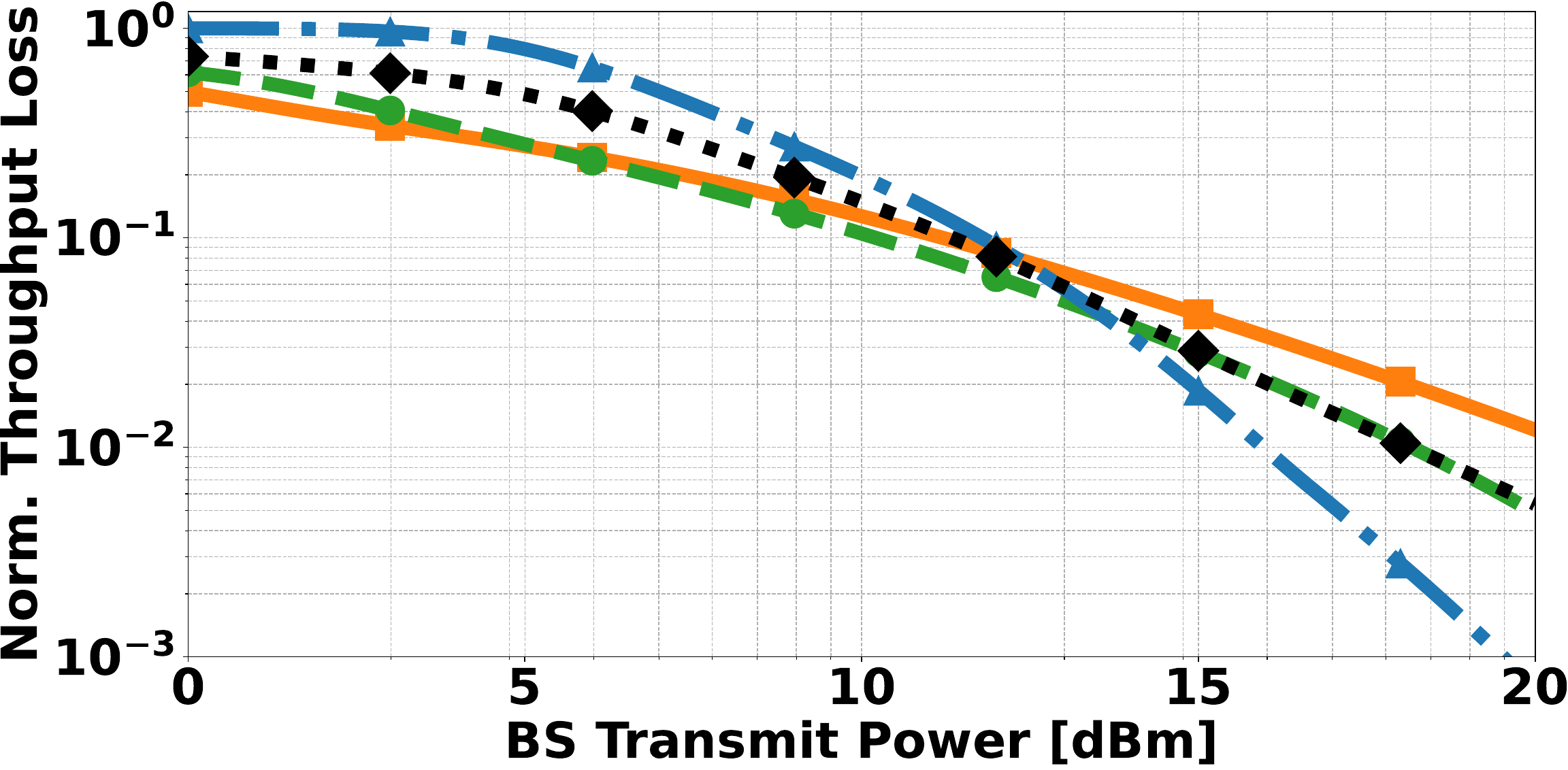}
        \centering (b) $\mathcal{L}_g$ vs. $P_t$
    \end{minipage}%
    \begin{minipage}{.33\linewidth}
    \includegraphics[width=6cm]{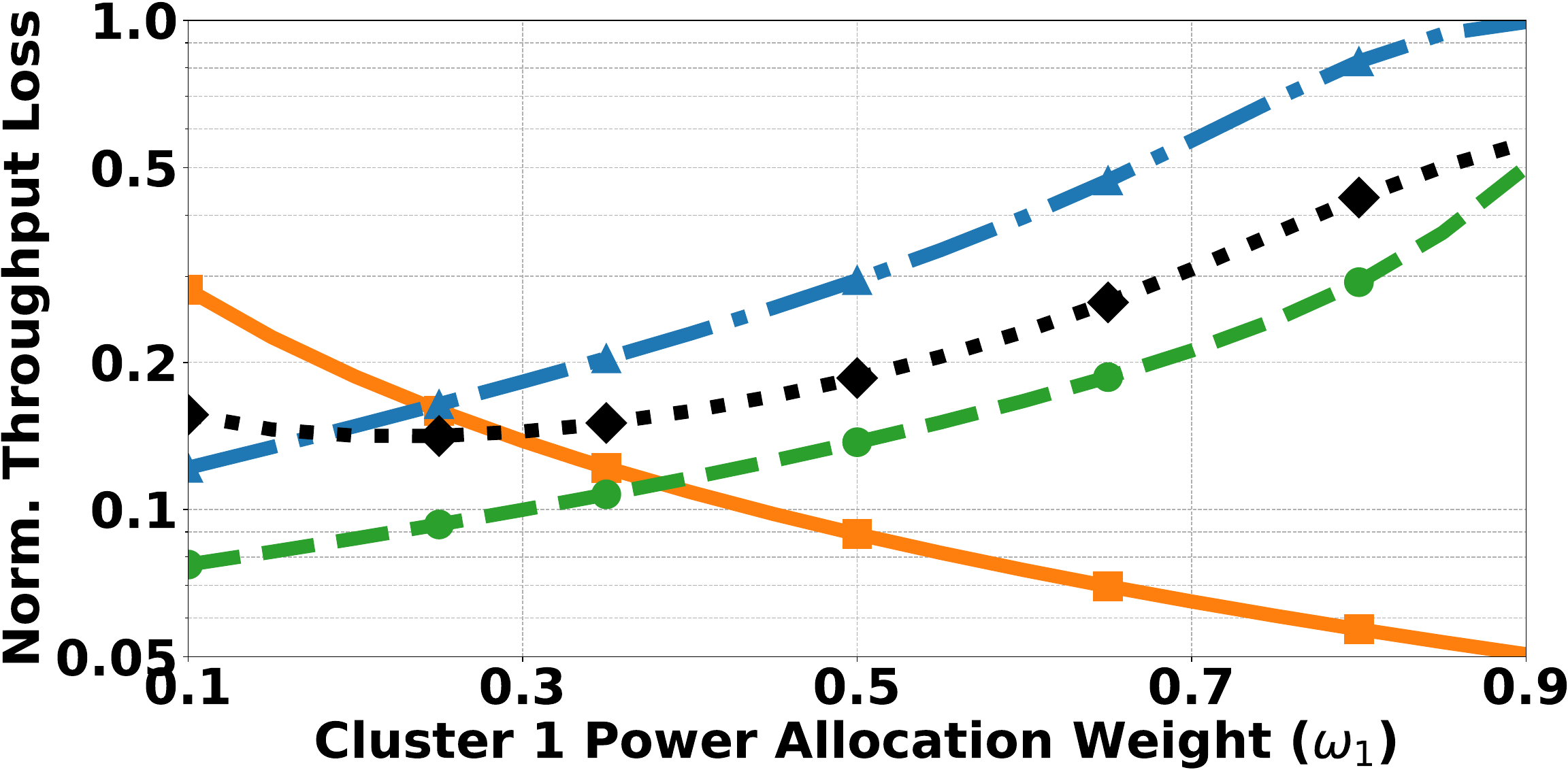}
        \centering (c) $\mathcal{L}_g$ vs. $\omega_1$
    \end{minipage}%
    \caption{Impact of (a) the RIS elements, (b) the BS transmit power, and (c) the power weight on $\mathcal{L}_g$ and $\mathcal{L}_{\text{net}} \triangleq \frac{\sum_{g} (\sum_{m_g}\hat{R}_{m_g})\mathcal{L}_g}{\sum_{g}\sum_{m_g}\hat{R}_{m_g}}$.}
\label{Fig:Normalized_throughput_loss}
\end{figure*}

\begin{figure*}[t]
    \centering
    \includegraphics[width=0.5\linewidth]{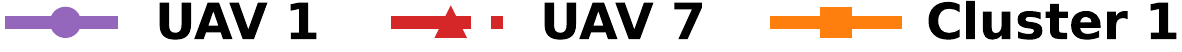}
    \begin{minipage}{.34\linewidth}
    \includegraphics[width=6cm]{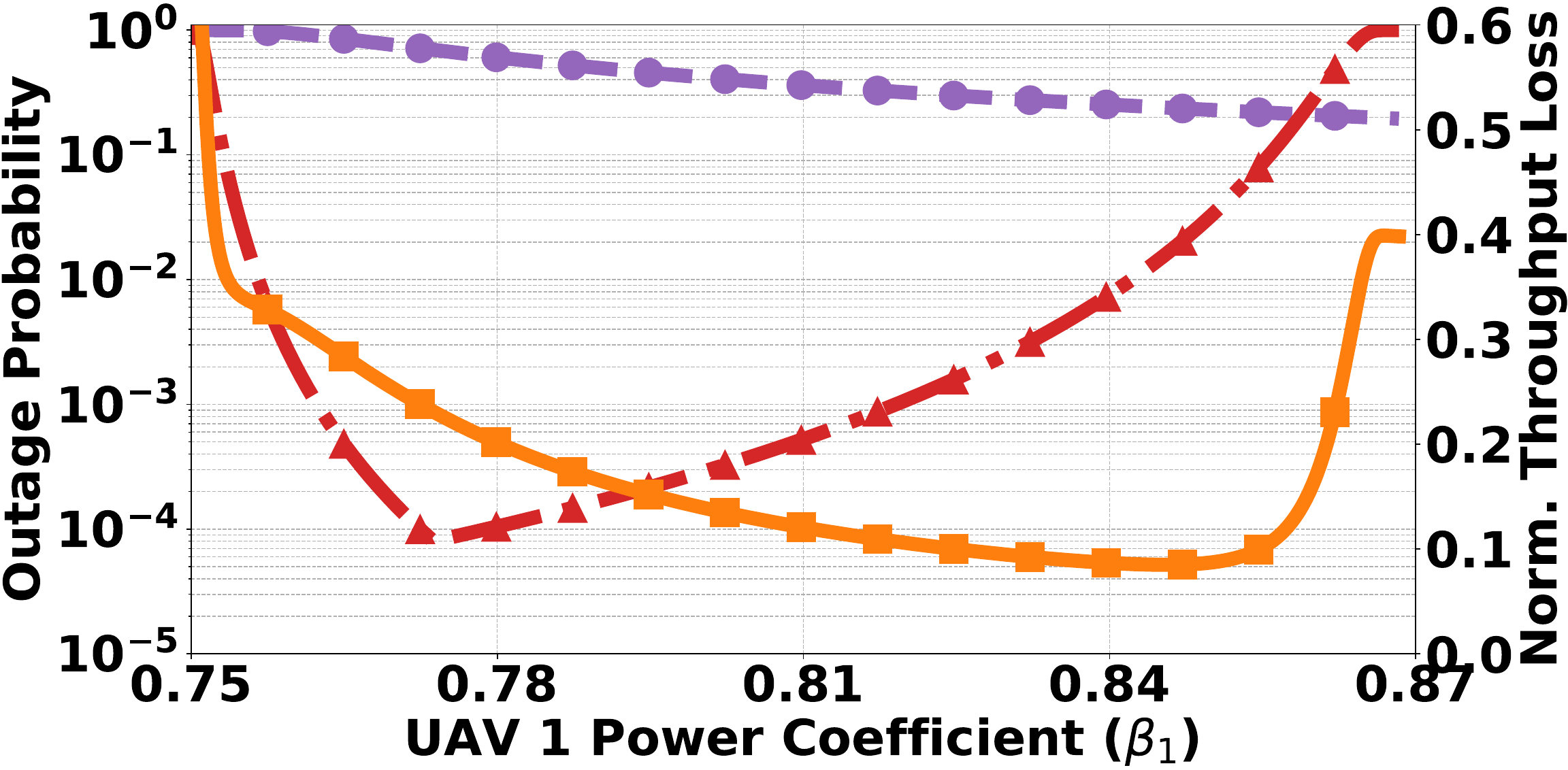}
        \centering (a) $\mathbb{P}_{m_g}^{out}$, $\mathcal{L}_1$ vs. $\beta_1$ 
    \end{minipage}%
    \begin{minipage}{.33\linewidth}
   \includegraphics[width=6cm]{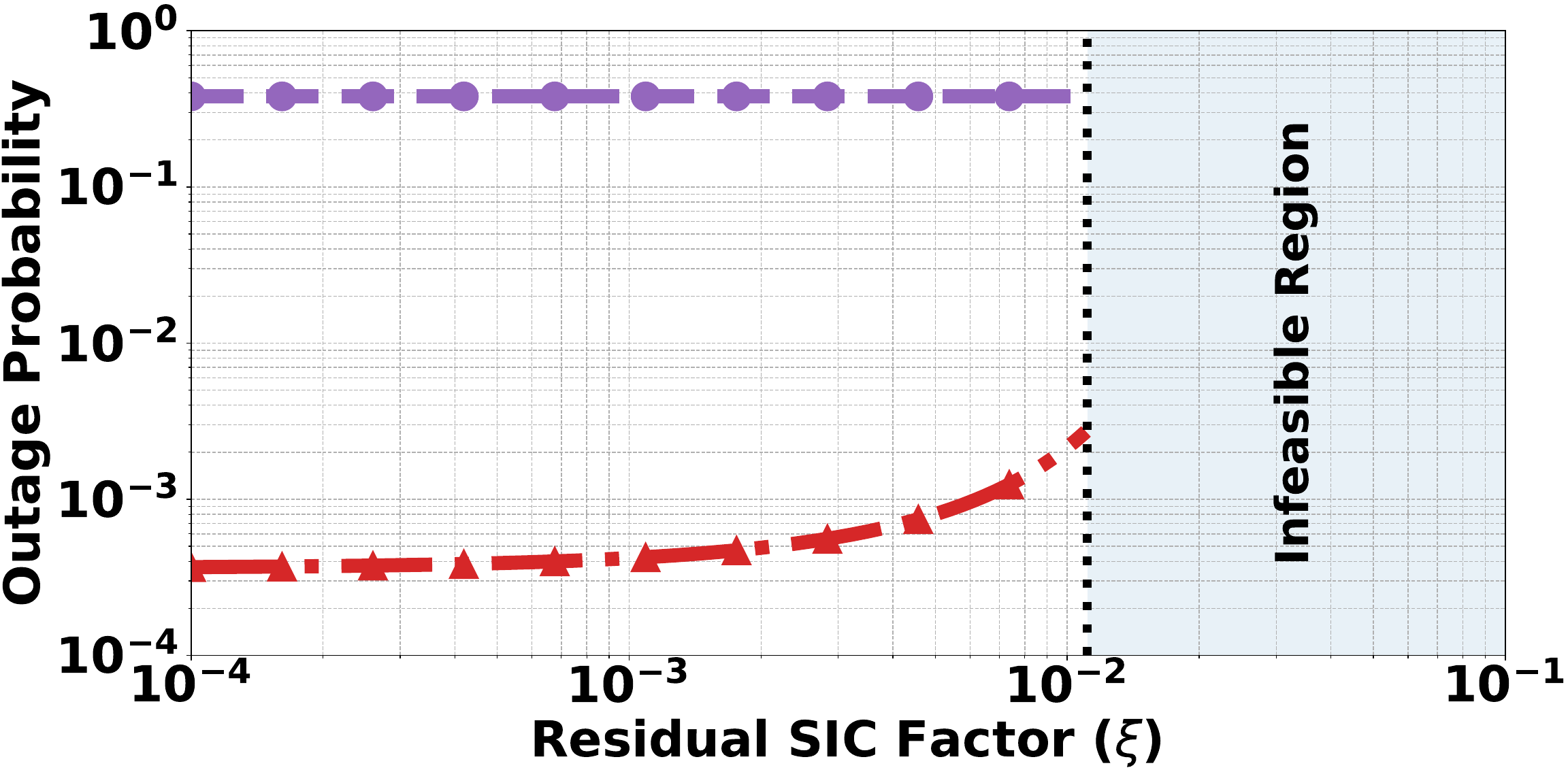}
        \centering (b) $\mathbb{P}_{m_g}^{out}$ vs. $\xi$
    \end{minipage}%
    \begin{minipage}{.33\linewidth}
    \includegraphics[width=6cm]{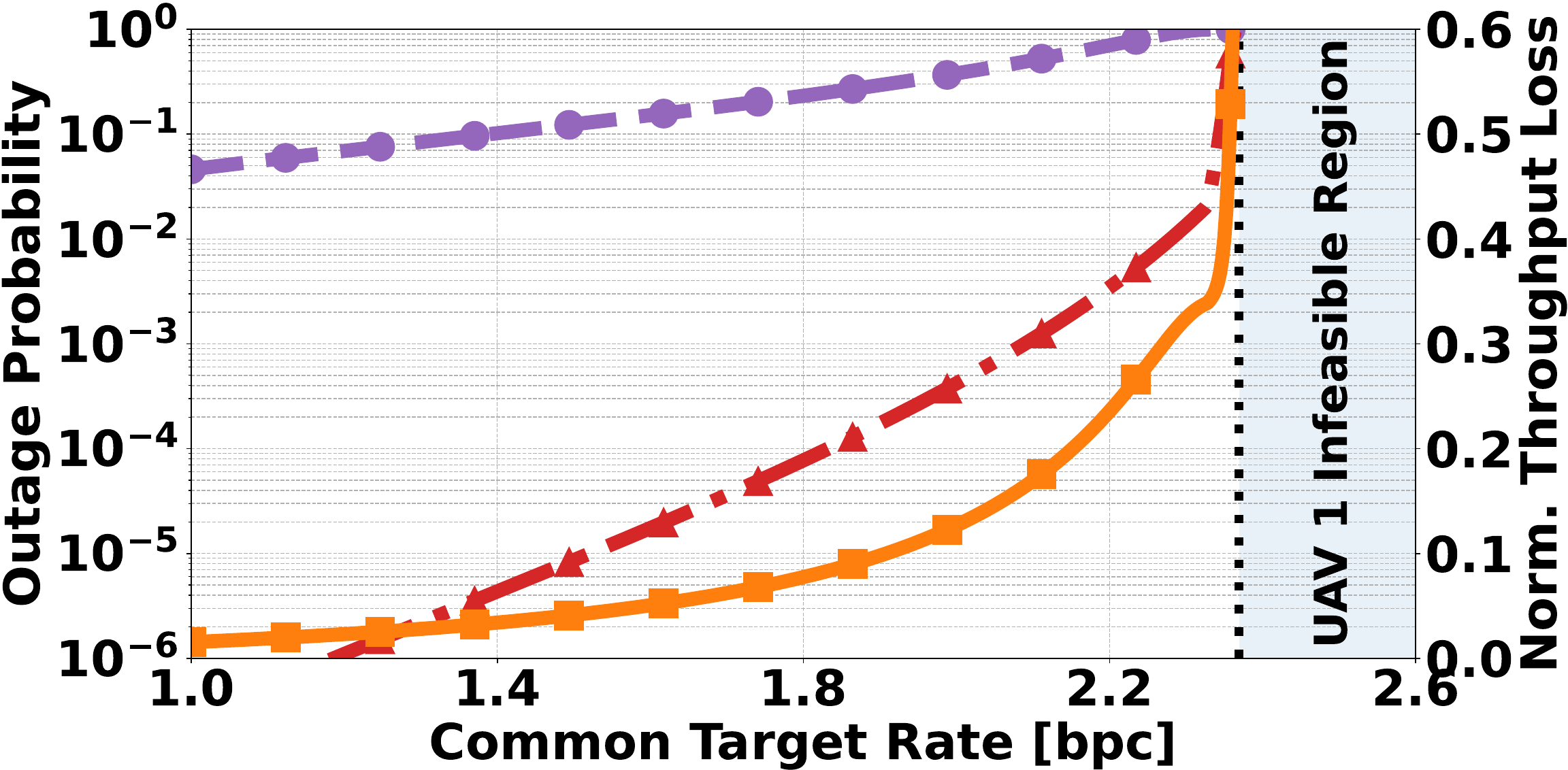}
        \centering (c) $\mathbb{P}_{m_g}^{out}$, $\mathcal{L}_1$ vs. $\hat{R}_{m_g}$ 
    \end{minipage}%
    \caption{Impact of (a) the UAV 1 power coefficient, (b) the residual SIC factor, and (c) the common target rate on $\mathbb{P}_{m_g}^{out}$ and $\mathcal{L}_1$.}
\label{Fig:outage_probability}
\end{figure*}

\subsection{Clustering Performance} \label{clustering_performance}
\subsubsection{Cluster Formation and RIS Assignment}
Fig.~\ref{Fig:3D_gris_clustering} illustrates GRIS clustering and RIS assignment for channel-only ($w_d=0$, $w_h=1$), balanced ($w_d=w_h=0.5$), and RIS-only ($w_d=1$, $w_h=0$) settings. The channel-only setting prioritizes the minimum intra-cluster channel gap, achieving $\delta_h^{(H)}=3.64$~dB but the highest RIS assignment cost among these settings, $C_d^{(H)}=2120.77$~m. The RIS-only setting reduces the assignment cost to $C_d^{(D)}=1529.39$~m at the expense of channel disparity, which decreases to $\delta_h^{(D)}=1.88$~dB. The balanced setting preserves the channel-only disparity of $3.64$~dB while reducing the assignment cost to $1943.04$~m, improving RIS proximity without sacrificing channel separation.

\subsubsection{Impact of $w_d$ and $w_h$}
Varying the weights from $0$ to $1$ in steps of $0.1$ yields only three distinct clustering configurations. Since UAV membership and RIS assignments are discrete, different weights can favor the same configuration, so each weight change need not produce a new clustering. Fig.~\ref{Fig:3D_gris_clustering} therefore presents the three representative channel-only, balanced, and RIS-only settings. Based on its trade-off between RIS assignment cost and channel disparity, the balanced setting and its clustering are adopted for the remaining results.

\begin{table}[t]
\centering
\small
\caption{Comparison of clustering methods (mean $\pm$ std.).}
\label{tab:clustering_comparison}
\begin{tabular}{lcc}
\toprule
Algorithm &
$C_d$ (m) &
$\delta_h$ (dB) \\
\midrule
Balanced Random &
$2219.75 \pm 187.07$ &
$1.20 \pm 0.89$ \\
Balanced K-means \cite{Katwe-2022-Dynamic} &
$1995.70 \pm 89.88$ &
$1.76 \pm 0.08$ \\
GRIS ($w_d=w_h=0.5$) &
$\mathbf{1943.04}$ &
$\mathbf{3.64}$ \\
\bottomrule
\end{tabular}
\end{table}

\subsubsection{Baseline Comparison}
Table~\ref{tab:clustering_comparison} compares GRIS with balanced random and balanced K-means clustering under approximately equal cluster sizes. Random clustering ignores spatial and channel information, whereas K-means exploits spatial proximity without explicitly promoting channel disparity. Baseline results are averaged over 1000 Monte Carlo trials, while GRIS yields identical results across the tested random seeds. GRIS achieves a lower RIS assignment cost and a larger minimum channel disparity than both baseline averages.

\begin{figure*}[t]
    \centering
    \begin{minipage}{.33\linewidth}
    \includegraphics[width=6cm]{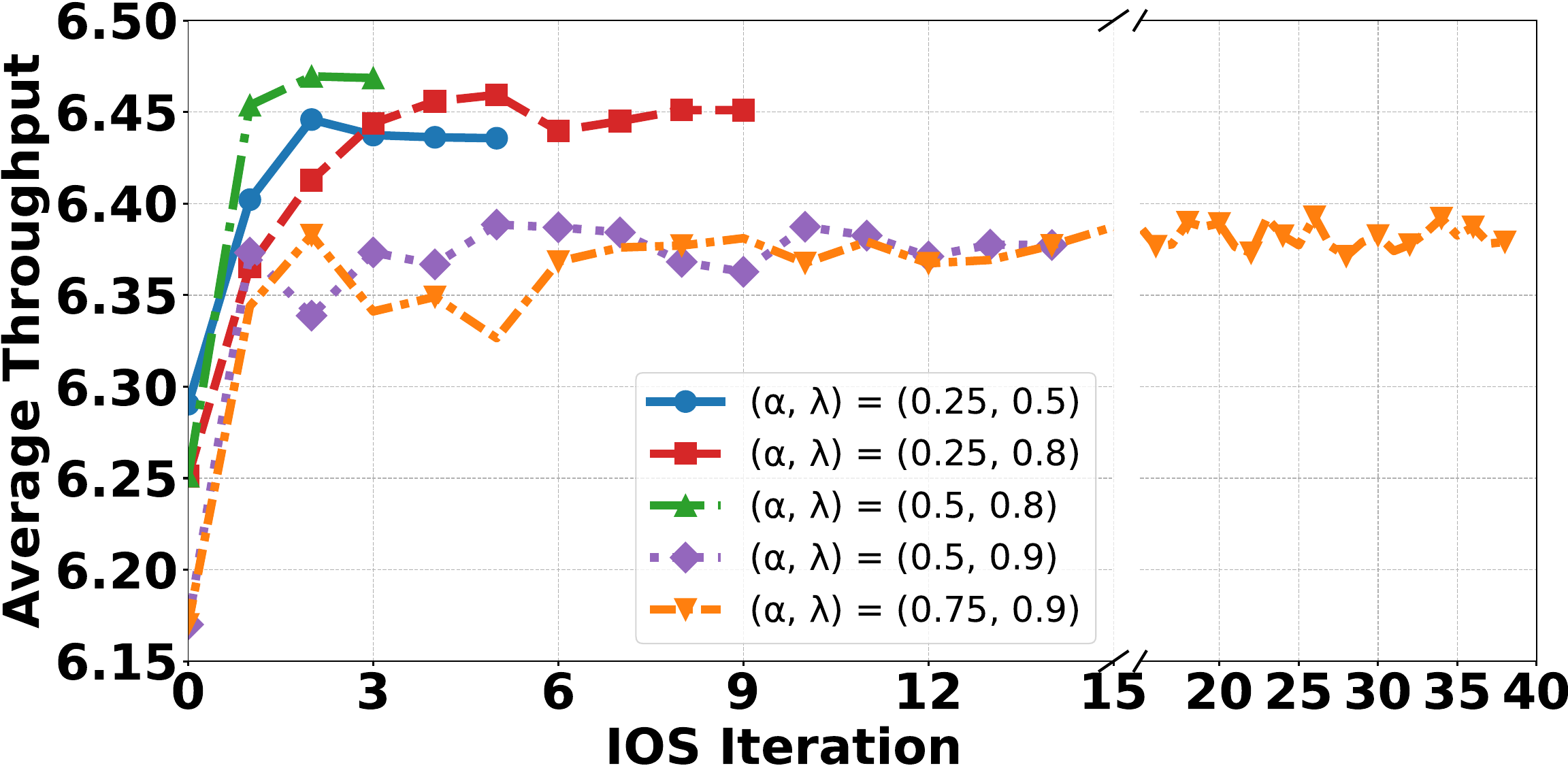}
        \centering (a) $\bar{R}$ vs. IOS Iteration
    \end{minipage}%
    \begin{minipage}{.34\linewidth}
   \includegraphics[width=6cm]{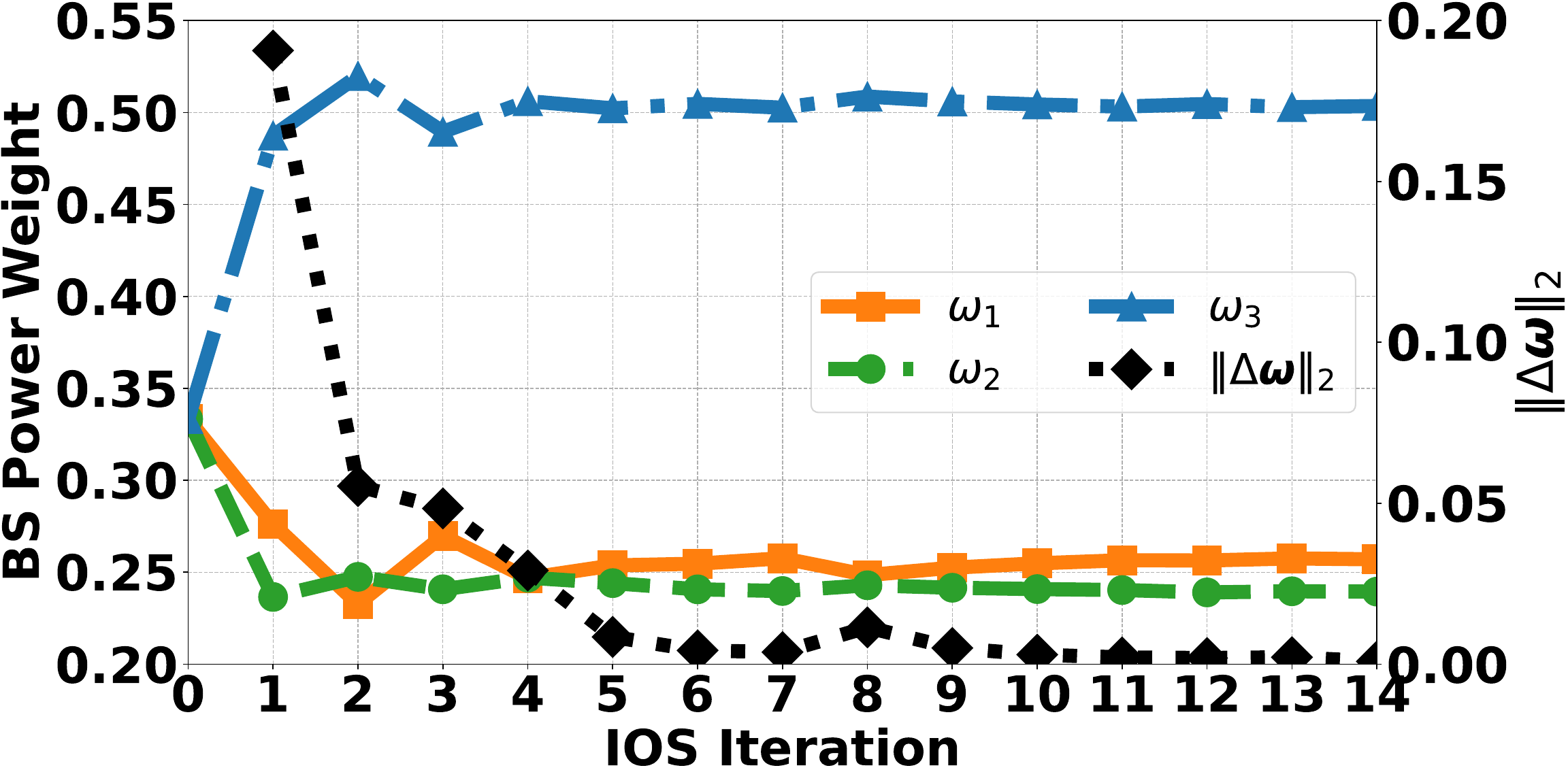}
        \centering (b) $\boldsymbol{\omega}$, $\|\Delta\boldsymbol{\omega}\|_2$ vs. IOS Iteration
    \end{minipage}%
    \begin{minipage}{.33\linewidth}
    \includegraphics[width=6cm]{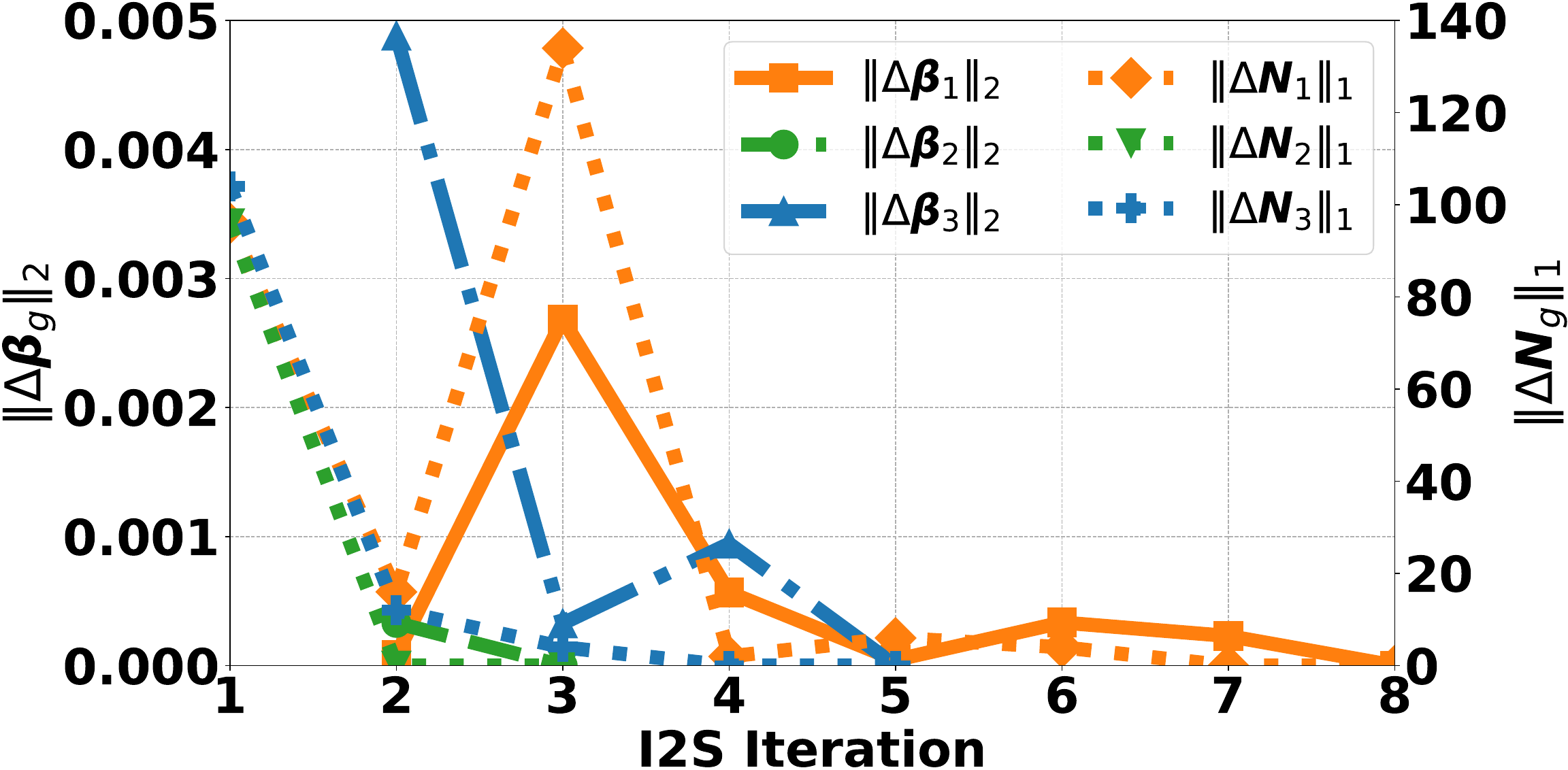}
        \centering (c) $\|\Delta\boldsymbol{\beta}_g\|_2$, $\|\Delta\boldsymbol{N}_g\|_1$ vs. I2S Iteration
    \end{minipage}%
    \caption{Convergence behavior of (a) the average throughput, (b) the BS power allocation and its update magnitude, and (c) the NOMA power coefficient and RIS element assignment update magnitudes.}
\label{Fig:convergence}
\end{figure*}

\subsection{Pre-Optimization Performance Analysis}
\subsubsection{Cluster-Level Performance}
Fig.~\ref{Fig:Normalized_throughput_loss} evaluates the normalized throughput loss, $\mathcal{L}_g \triangleq(\sum_{m_g}\hat{R}_{m_g}-R_g)/(\sum_{m_g}\hat{R}_{m_g})$, under different parameters. Normalization by the aggregate target rate facilitates comparisons across clusters of different sizes. Clusters 1 and 2, shown in orange and green, respectively, in Fig.~\ref{Fig:3D_gris_clustering}(b), each contain three UAVs with fixed $\boldsymbol{\beta}_1=\boldsymbol{\beta}_2\approx[0.8065, 0.1613, 0.0323]$, whereas Cluster 3 contains four UAVs with $\boldsymbol{\beta}_3\approx[0.8013, 0.1603, 0.0321, 0.0064]$. Unless varied, BS power is equally allocated among clusters, and each cluster’s RIS elements are equally divided among its UAVs.

Fig.~\ref{Fig:Normalized_throughput_loss}(a) illustrates the impact of the total number of RIS elements assigned to each cluster. Increasing the element budget reduces $\mathcal{L}_g$ for all clusters by strengthening the composite channel. Cluster 3 exhibits the largest reduction, from $0.362$ to $0.196$, but retains the highest $\mathcal{L}_g$ throughout the range. Clusters 1 and 2 show smaller reductions, reaching $0.127$ and $0.105$, respectively, at $144$ elements. The network-level loss decreases monotonically from $0.227$ to $0.148$, demonstrating the benefit of additional RIS resources. Fig.~\ref{Fig:Normalized_throughput_loss}(b) shows that increasing the BS transmit power reduces $\mathcal{L}_g$ for all clusters. Cluster 3 initially has the highest loss but exhibits the largest reduction, achieving the lowest $\mathcal{L}_g$ at higher powers. At $20$~dBm, $\mathcal{L}_g$ reaches $0.0121$, $0.0049$, and $0.0007$ for Clusters 1, 2, and 3, respectively. The network-level loss also decreases monotonically, reflecting the overall throughput improvement. Fig.~\ref{Fig:Normalized_throughput_loss}(c) examines the impact of $\omega_1$ with $\omega_2=\omega_3=(1-\omega_1)/2$. Increasing $\omega_1$ reduces $\mathcal{L}_1$ by allocating more power to Cluster 1, while increasing the losses of Clusters 2 and 3. The network-level loss reaches its lowest value of $0.1414$ at $\omega_1=0.25$, illustrating the trade-off among clusters and motivating the optimization of inter-cluster BS power allocation.

\subsubsection{UAV-Level Performance}
Fig.~\ref{Fig:outage_probability}(a) shows the impact of $\beta_1$ on $\mathbb{P}_{m_g}^{out}$ and $\mathcal{L}_1$ for Cluster 1, with $\beta_3\approx0.0323$ and $\beta_2=1-\beta_1-\beta_3$. Increasing $\beta_1$ reduces UAV 1's outage, whereas UAV 7's outage first decreases and then increases, reflecting the trade-off between easier decoding of UAV 1 for SIC and less power allocated to UAV 7. $\max_{m_g}\mathbb{P}_{m_g}^{out}$ is minimized near $\beta_1=0.8588$, whereas $\mathcal{L}_1$ reaches its minimum of $0.0850$ near $\beta_1=0.8455$, illustrating distinct allocations for reliability and efficiency. UAV 3's negligible outage is not plotted, but its contribution is included in $\mathcal{L}_1$. Fig.~\ref{Fig:outage_probability}(b) shows the impact of $\xi$ on $\mathbb{P}_{m_g}^{out}$ for Cluster 1. UAV 1's outage remains unchanged because it performs no SIC, whereas UAV 7's outage increases with residual interference from cancelling UAV 1's signal. The cluster's SIC feasibility limit is $\xi_{\max}\approx0.0111$, determined by UAV 3's constraint. The vertical dotted line and shaded region indicate this boundary and the infeasible range, respectively, highlighting the importance of controlling residual SIC interference. Fig.~\ref{Fig:outage_probability}(c) shows the impact of $\hat{R}_{m_g}$, same for all UAVs, on $\mathbb{P}_{m_g}^{out}$ and $\mathcal{L}_1$ for Cluster 1. Higher $\hat{R}_{m_g}$ requires higher SINR thresholds, increasing UAV 1's outage. UAV 7 maintains low outage over much of the feasible range but experiences a sharp increase near the feasibility boundary, accompanied by a rapid increase in $\mathcal{L}_1$. This boundary occurs at $2.369$~bpc, determined by UAV 1's decoding feasibility constraint. These results illustrate the trade-off between higher $\hat{R}_{m_g}$ and reliable NOMA decoding.

\begin{figure}
    \centering
    \includegraphics[width=0.8\linewidth]{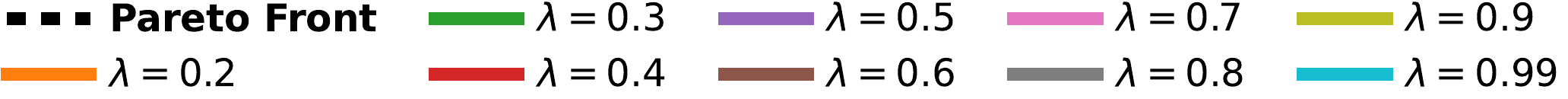}
    \includegraphics[width=0.9\linewidth]{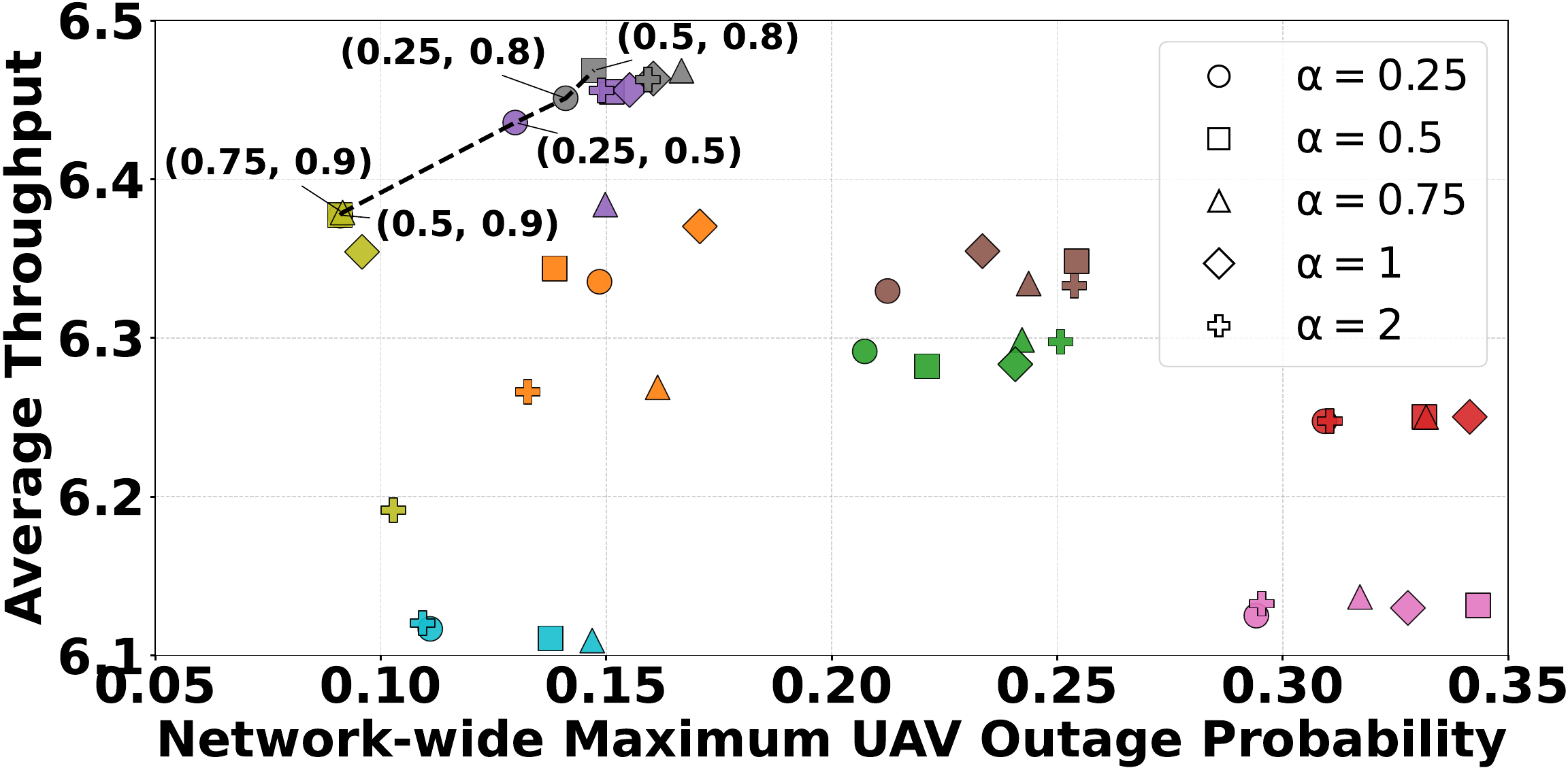}
    \caption{$\bar{R}$ vs. $\max_{g,m_g}\mathbb{P}_{m_g}^{out}$ across different $(\alpha,\lambda)$ configurations, highlighting the Pareto front and the efficiency-reliability trade-off.}
    \label{fig:pareto}
\end{figure}

\subsection{Performance Evaluation of the Proposed Optimization}

\begin{table*}[t]
\centering
\caption{Optimized IOS-I2S results for the selected configuration
$(\alpha,\lambda)=(0.5,0.9)$.}
\label{tab:selected_configuration}
\resizebox{\textwidth}{!}{%
\begin{tabular}{c c c c c c c}
\hline
$g$
& $\omega_g^\star$
& $\boldsymbol{\beta}_g^\star$
& $\boldsymbol{N}_g^\star$
& $\mathbb{P}_{m_g}^{out}$
& $J_g$
& $R_g$ (bpc)
\\
\hline

1
& $0.2570$
& $[0.9671,\,0.0299,\,0.0030]$
& $[144,\,0,\,0]$
& $[0.09095,\,0.08297,\,0.00246]$
& $0.8236$
& $5.6472$
\\

2
& $0.2395$
& $[0.9586,\,0.0383,\,0.0031]$
& $[73,\,71,\,0]$
& $[0.07527,\,0.07475,\,8.93\times10^{-16}]$
& $0.4356$
& $5.7$
\\

3
& $0.5035$
& $[0.8036,\,0.1661,\,0.0254,\,0.0049]$
& $[50,\,79,\,15,\,0]$
& $[0.02698,\,0.02698,\,0.02672,\,0.02684]$
& $1.0$
& $7.7849$
\\

\hline
\end{tabular}%
}
\end{table*}

\subsubsection{Hyperparameter Sensitivity and Convergence Analysis}
Fig.~\ref{fig:pareto} illustrates the efficiency-reliability trade-off obtained by jointly varying the IOS step size $\alpha$ and the I2S refinement factor $\lambda$, where reliability is characterized by the network-wide maximum UAV outage probability, i.e., $\max_g\max_{m_g}\mathbb{P}_{m_g}^{out}$. The Pareto front consists of five non-dominated operating points, spanning from reliability-oriented to efficiency-oriented configurations. At the reliability end, $(\alpha,\lambda)=(0.5,0.9)$ achieves the lowest maximum UAV outage of $0.091$ while maintaining a high average throughput of $6.377$ bpc, whereas $(0.5,0.8)$ achieves the highest average throughput of $6.469$ bpc at a higher maximum UAV outage of $0.147$. The remaining Pareto-optimal configurations, $(0.75,0.9)$, $(0.25,0.5)$, and $(0.25,0.8)$, provide intermediate efficiency-reliability trade-offs. In terms of Jain's fairness, which is defined as \cite{Sediq-2013-Optimal}: 
\begin{equation*}
J_g(\omega_g,\boldsymbol{\beta}_g,\boldsymbol{N}_g^k)=
\frac{\big(\sum_{m_g=1}^{M_g}
\log \frac{1}{\mathbb{P}_{m_g}^{out}(\omega_g,\boldsymbol{\beta}_g,N_{m_g}^k)}\big)^2}
{M_g\sum_{m_g=1}^{M_g}
\big(\log \frac{1}{\mathbb{P}_{m_g}^{out}(\omega_g,\boldsymbol{\beta}_g,N_{m_g}^k)}\big)^2 },
\end{equation*}
$(\alpha,\lambda)=(2,0.9)$ achieves the highest average fairness of $0.998$ across the clusters, with cluster-wise values of $J_1=0.995$, $J_2\approx1.000$, and $J_3\approx1.000$; however, this setting yields a lower average throughput of $6.191$ bpc and a maximum UAV outage above $0.1$. Among the Pareto-optimal configurations, $(0.5,0.9)$ is therefore selected for the remaining evaluations because it maintains the maximum UAV outage below $0.1$, preserves a high average throughput, and converges considerably faster than the closely performing $(0.75,0.9)$ configuration.

Fig.~\ref{Fig:convergence}(a) investigates the convergence behavior of the five Pareto-optimal $(\alpha,\lambda)$ configurations in terms of the average throughput. All configurations exhibit a rapid throughput improvement during the first few iterations, followed by relatively small variations as the IOS updates approach convergence. The configurations $(0.25,0.5)$, $(0.25,0.8)$, and $(0.5,0.8)$ converge within only a few iterations and achieve comparatively high average throughput values of approximately $6.44$-$6.47$ bpc, whereas the selected configuration $(0.5,0.9)$ converges after 14 iterations to approximately $6.377$ bpc. In contrast, $(0.75,0.9)$ requires substantially more iterations, converging after 38 iterations to approximately $6.379$ bpc, while exhibiting small oscillations around its steady-state throughput.

Fig.~\ref{Fig:convergence}(b) illustrates the evolution of the BS power weights and the corresponding update magnitude. Starting from the equal allocation $\omega_1=\omega_2=\omega_3=1/3$, IOS initially reallocates a larger fraction of the BS power to Cluster~3, increasing $\omega_3$ to approximately $0.49$ in the first iteration and above $0.50$ thereafter, while reducing $\omega_1$ and $\omega_2$. The power weights then show progressively smaller adjustments and converge after 14 IOS iterations to approximately $\boldsymbol{\omega}^{\star}=[0.257,0.239,0.504]$. This stabilization is also reflected by $\|\Delta\boldsymbol{\omega}\|_2$, which decreases sharply after the first update and remains close to zero in the later iterations. Overall, the results show that IOS initially shifts power toward the cluster providing the largest marginal throughput improvement and then refines the BS power allocation through diminishing updates until convergence.

Fig.~\ref{Fig:convergence}(c) shows I2S convergence at the final BS power allocation $\boldsymbol{\omega}^{\star}$ by showing the update magnitudes of the NOMA power coefficients and RIS element assignment for all clusters. The power allocation updates, measured by $\|\Delta\boldsymbol{\beta}_g\|_2$, decrease toward zero as the I2S iterations proceed, indicating stabilization of the optimized coefficients. Similarly, the element assignment updates, measured by $\|\Delta\boldsymbol{N}_g\|_1$, exhibit relatively large changes during the initial iterations as elements are redistributed among the UAVs, followed by rapid reduction to zero once the assignments stabilize. Cluster 2 converges in a few iterations, whereas Clusters 1 and 3 require additional updates before both decision variables become stationary.

Finally, Table~\ref{tab:selected_configuration} summarizes the optimized resource allocation obtained for the selected configuration $(\alpha,\lambda)=(0.5,0.9)$. The final IOS solution allocates the BS power weights as $\boldsymbol{\omega}^{\star}=[0.2570,\,0.2395,\,0.5035]$, with the largest fraction allocated to Cluster~3. The corresponding I2S solutions jointly determine the NOMA power coefficients and RIS element assignments for each cluster. The resulting cluster throughputs are $5.6472$, $5.7$, and $7.7849$ bpc for Clusters~1-3, respectively, while the maximum UAV outage remains below $0.1$ across all clusters. These results provide the final optimized operating point used in the subsequent performance comparisons.

\begin{table}[t]
\centering
\caption{Performance comparison with baseline schemes.}
\label{tab:baseline}
\begin{tabular}{l c c}
\hline
Scheme
& $\bar{R}$ (bpc)
& $\max_{g,m_g}\mathbb{P}_{m_g}^{out}$
\\
\hline
\textbf{IOS-I2S (Proposed)}         & \textbf{6.3774} & \textbf{0.0909} \\
OMA \cite{Mu-2021-Intelligent}              & 1.9960 & 0.0176 \\
NoRIS \cite{Mu-2021-Intelligent}             & 5.4930 & 0.4030 \\
ERA  & 5.6795 & 0.4790 \\
EOA \cite{Xing-2018-Optimal}               & 6.5212 & 0.1067 \\
ROA \cite{Timotheou-2015-Fairness}               & 6.3618 & 0.0752 \\
\hline
\end{tabular}
\end{table}

\subsubsection{Performance Comparison with Baseline Schemes}
To evaluate the effectiveness of the proposed IOS-I2S framework, we compare it with five baseline schemes adapted to our system model and resource constraints, isolating the contributions of multiple access, RIS assistance, resource optimization, and optimization objectives. The first baseline, denoted by \textit{orthogonal multiple access (OMA)} \cite{Mu-2021-Intelligent}, replaces intra-cluster NOMA with OMA. The second baseline, denoted by \textit{NOMA without RIS (NoRIS)} \cite{Mu-2021-Intelligent}, removes RIS assistance. The third baseline, denoted by \textit{equal resource allocation (ERA)}, uses $\boldsymbol{\omega}=[1/3,1/3,1/3]$, uniform RIS element assignment $N_{m_g}=144/M_g$, and fixed $\boldsymbol{\beta}_1\approx\boldsymbol{\beta}_2\approx[0.8065,\,0.1613,\,0.0323]$ and $\boldsymbol{\beta}_3\approx[0.8013,\,0.1603,\,0.0321,\,0.0064]$. The fourth baseline, denoted by \textit{efficiency-oriented allocation (EOA)} \cite{Xing-2018-Optimal}, optimizes the decision variables solely to maximize the average throughput $\bar{R}$. The fifth and final baseline, denoted by \textit{reliability-oriented allocation (ROA)} \cite{Timotheou-2015-Fairness}, optimizes the decision variables solely to minimize the maximum UAV outage both within each cluster and across all clusters.

Table~\ref{tab:baseline} compares IOS-I2S with the baseline schemes in terms of average throughput and network-wide maximum UAV outage probability. Compared with NoRIS and ERA, IOS-I2S increases the average throughput by approximately 16.1\% and 12.3\%, respectively, while reducing the maximum outage by 77.4\% and 81.0\%, highlighting the benefits of RIS assistance and optimized resource allocation. OMA achieves the lowest maximum outage of 0.0176, but its average throughput is limited to 1.9960~bpc, while IOS-I2S provides approximately 3.20 times this throughput at a higher outage level. OMA benefits from eliminating intra-cluster interference and allocating full cluster power to each active UAV, while sharing transmission time among UAVs limits its efficiency. EOA achieves the highest throughput of 6.5212~bpc, whereas IOS-I2S reduces the maximum outage from 0.1067 to 0.0909, corresponding to a 14.8\% reduction with a 2.2\% throughput decrease. EOA prioritizes efficiency by selecting allocations based on average throughput, without a reliability objective. ROA prioritizes reliability, achieving a lower maximum outage of 0.0752 while maintaining a throughput of 6.3618~bpc, only 0.0156~bpc below IOS-I2S. Thus, IOS-I2S yields an intermediate efficiency-reliability operating point between EOA and ROA, while improving both metrics over NoRIS and ERA.

\begin{table}[t]
\centering
\caption{Ablation study over all combinations of $\boldsymbol{\omega}$, $\boldsymbol{\beta}_g$, and $\boldsymbol{N}_g$.}
\label{tab:ablation}
\begin{tabular}{c c c c c}
\hline
$\boldsymbol{\omega}^{\star}$
& $\boldsymbol{\beta}_g^{\star}$
& $\boldsymbol{N}_g^{\star}$
& $\bar{R}$ (bpc)
& $\max_{g,m_g}\mathbb{P}_{m_g}^{out}$
\\
\hline
\checkmark & \checkmark & \checkmark & $\textbf{6.3774}$ & $\textbf{0.0909}$ \\
$\times$   & \checkmark & \checkmark & $6.1701$ & $0.1055$ \\
\checkmark & $\times$   & \checkmark & $6.1093$ & $0.4085$ \\
\checkmark & \checkmark & $\times$   & $6.1798$ & $0.1586$ \\
$\times$   & $\times$   & \checkmark & $5.9751$ & $0.3147$ \\
$\times$   & \checkmark & $\times$   & $5.9951$ & $0.2348$ \\
\checkmark & $\times$   & $\times$   & $5.8034$ & $0.4890$ \\
$\times$   & $\times$   & $\times$   & $5.6795$ & $0.4790$ \\
\hline
\end{tabular}%
\end{table}

\subsubsection{Decision-Variable Ablation Study}
Table~\ref{tab:ablation} evaluates the contribution of the three optimized decision variables. Whenever a decision variable is fixed, its value follows the ERA settings. Joint optimization of all three variables achieves the highest average throughput of $6.3774$ bpc while also yielding the lowest network-wide maximum UAV outage of $0.0909$. Disabling only $\boldsymbol{\omega}$ reduces the average throughput to $6.1701$ bpc and increases the maximum outage to $0.1055$, whereas disabling $\boldsymbol{N}_g$ increases the maximum outage to $0.1586$. One of the largest reliability losses occurs when $\boldsymbol{\beta}_g$ is not optimized, increasing the maximum UAV outage to $0.4085$ even when $\boldsymbol{\omega}$ and $\boldsymbol{N}_g$ remain optimized. The ablation cases show that optimizing only a subset of the decision variables cannot recover the performance of the complete framework.

\section{Conclusion} \label{conclusion}
In this paper, we investigated RIS-assisted clustered BS-UAV downlink NOMA under imperfect SIC. We developed an analytical framework for the outage performance of direct, RIS-only, and composite links under Nakagami-$m$ and double Nakagami-$m$ fading, and introduced GRIS clustering to jointly account for RIS proximity and channel disparity. We formulated a bilevel efficiency-reliability problem and developed IOS-I2S to jointly optimize inter-cluster BS power allocation, intra-cluster NOMA power coefficients, and RIS element assignment. Numerical comparisons showed that GRIS improved both clustering metrics over the baseline averages, while IOS-I2S improved both efficiency and reliability over NoRIS and ERA and provided an intermediate operating point between EOA and ROA. Ablation results highlighted the importance of jointly optimizing all three variables. Future extensions include mobility-aware clustering to update UAV groups as their positions change and learning-based online resource optimization that adapts to dynamic, non-stationary channels.

\bibliographystyle{IEEEtran}
\bibliography{ref}

@article{Yang-2020-Performance,
	title        = {On the Performance of {RIS}-Assisted Dual-Hop {UAV} Communication Systems},
	author       = {Yang, Liang and Meng, Fanxu and Zhang, Jiayi and Hasna, Mazen O. and Renzo, Marco Di},
	year         = 2020,
	journal      = {IEEE Transactions on Vehicular Technology}
}

@article{Yang-2022-Performance,
	title        = {Performance Analysis of {RIS}-Assisted {UAV} Communication Systems},
	author       = {Yang, Liang and Li, Pinhui and Meng, Fanxu and Yu, Siyuan},
	year         = 2022,
	journal      = {IEEE Trans. Veh. Technol.}
}

@article{Abualhayja-2024-Exploiting,
	title        = {Exploiting Multi-Hop {RIS}-Assisted {UAV} Communications: Performance Analysis},
	author       = {Abualhayja’a, Mohammad and Centeno, Anthony and Mohjazi, Lina and Butt, M. Majid and Sehier, Philippe and Imran, Muhammad Ali},
	year         = 2024,
	journal      = {IEEE Communications Letters}
}

@article{Tegos-2022-Distribution,
	title        = {On the Distribution of the Sum of Double-Nakagami-$m$ Random Vectors and Application in Randomly Reconfigurable Surfaces},
	author       = {Tegos, Sotiris A. and Tyrovolas, Dimitrios and Diamantoulakis, Panagiotis D. and Liaskos, Christos K. and Karagiannidis, George K.},
	year         = 2022,
	journal      = {IEEE Transactions on Vehicular Technology}
}

@article{Kim-2019-Impact,
	title        = {Impact of an Interfering Node on Unmanned Aerial Vehicle Communications},
	author       = {Kim, Minsu and Lee, Jemin},
	year         = 2019,
	journal      = {IEEE Trans. on Vehicular Technology}
}

@article{Zhang-2022-Joint,
	title        = {Joint Optimization of {STAR-RIS} Assisted {UAV} Communication Systems},
	author       = {Zhang, Qin and Zhao, Yang and Li, Hai and Hou, Shujuan and Song, Zhengyu},
	year         = 2022,
	journal      = {IEEE Wireless Communications Letters}
}

@article{Dorra-2011-SIR,
	title        = {{SIR} distribution analysis in cellular networks considering the joint impact of path-loss, shadowing and fast fading},
	author       = {Dorra Ben Cheikh and Jean-Marc Kelif and Marceau Coupechoux and Philippe Godlewski},
	year         = 2011,
	journal      = {EURASIP Journal on Wireless Communications and Networking}
}

@book{David-2003-Order,
	title        = {Order Statistics},
	author       = {David, H. A. and Nagaraja, H. N.},
	year         = 2003,
	publisher    = {Wiley},
	edition      = {3rd}
}

@article{Alqahtani-2021-Performance,
	title        = {Performance Analysis for Downlink {NOMA} Over $\alpha$-$\mu$ Generalized Fading Channels},
	author       = {Alqahtani, Adel and Alsusa, Emad and Al-Dweik, Arafa and Al-Jarrah, Mohammad},
	year         = 2021,
	journal      = {IEEE Transactions on Vehicular Technology}
}

@article{Channel_Ghazikor_2024,
	title        = {Channel-Aware Distributed Transmission Control and Video Streaming in {UAV} Networks},
	author       = {Ghazikor, Masoud and Roach, Keenan  and Cheung, Kenny and Hashemi, Morteza},
	year         = {2026},
	journal      = {IEEE Transactions on Communications}
}

@misc{FAA_UAV_2026,
  author       = {FAA},
  title        = {{FAA} By the Numbers},
  year         = {2026},
  url          = {https://www.faa.gov/node/26},
  note         = {Released: June., 2026}
}

@article{Analysis_Ni_2023,
	title        = {Analysis of {RIS}-Aided Communications Over Nakagami-$m$ Fading Channels},
	author       = {Ni, Yiyang and Zhao, Haitao and Liu, Yaxuan and Wang, Jue and Gui, Guan and Zhang, Hui},
	year         = 2023,
	journal      = {IEEE Transactions on Vehicular Technology}
}

@article{Chen-2024-Optimizing,
	title        = {Optimizing {NOMA} Transmissions to Advance Federated Learning in Vehicular Networks},
	author       = {Chen, Ziru and Ni, Zhou and Guan, Peiyuan and Wang, Lu and Cai, Lin X. and Hashemi, Morteza and Li, Zongzhi},
	year         = 2024,
	journal      = {Proc. IEEE Global Communications Conf.}
}

@article{Tang-2025-Throughput,
	title        = {Throughput Maximization in {NOMA} Enhanced {RIS}-Assisted Multi-{UAV} Networks: A Deep Reinforcement Learning Approach},
	author       = {Tang, Runzhi and Wang, Junxuan and Zhang, Yanyan and Jiang, Fan and Zhang, Xuewei and Du, Jianbo},
	year         = 2025,
	journal      = {IEEE Trans. Veh. Technol.}
}

@article{Sobhi-2025-Efficient,
	title        = {Efficient Optimization in {RIS}-Assisted {UAV} System Using Deep Reinforcement Learning for mmWave-{NOMA} {6G} Communications},
	author       = {Sobhi-Givi, Sima and Nouri, Mahdi and Shayesteh, Mahrokh G. and Behroozi, Hamid and Kwon, Hyun Han and Piran, Md. Jalil},
	year         = 2025,
	journal      = {IEEE Internet Things Journal}
}

@article{Li-2024-Performance,
	title        = {Performance Analysis of Double Reconfigurable Intelligent Surfaces Assisted {NOMA} Networks},
	author       = {Li, Xuehua and Lian, Xuanhao and Yue, Xinwei and Lu, Zhiping and Huang, Chongwen and Hou, Tianwei},
	year         = 2024,
	journal      = {IEEE Transactions on Vehicular Technology}
}

@article{Jafarkhani-2024-Modulation,
	title        = {Modulation and Coding for {NOMA} and {RSMA}},
	author       = {Jafarkhani, Hamid and Maleki, Hossein and Vaezi, Mojtaba},
	year         = 2024,
	journal      = {Proceedings of the IEEE}
}

@article{Zeng-2017-Capacity,
	title        = {Capacity Comparison Between {MIMO-NOMA} and {MIMO-OMA} With Multiple Users in a Cluster},
	author       = {Zeng, Ming and Yadav, Animesh and Dobre, Octavia A. and Tsiropoulos, Georgios I. and Poor, H. Vincent},
	year         = 2017,
	journal      = {IEEE J. Sel. Areas Commun.}
}

@article{Mahmoudi-2022-User,
	title        = {User Clustering and Resource Allocation in Hybrid {NOMA-OMA} Systems Under Nakagami-m Fading},
	author       = {Mahmoudi, Ali and Abolhassani, Bahman and Razavizadeh, S. Mohammad and Nguyen, Ha H.},
	year         = 2022,
	journal      = {IEEE Access}
}

@article{Sediq-2013-Optimal,
	title        = {Optimal Tradeoff Between Sum-Rate Efficiency and Jain's Fairness Index in Resource Allocation},
	author       = {Sediq, Akram Bin and Gohary, Ramy H. and Schoenen, Rainer and Yanikomeroglu, Halim},
	year         = 2013,
	journal      = {IEEE Trans. Wireless Communications}
}

@article{Chintareddy-2025-Federated,
	title        = {Federated Learning-Based Collaborative Wideband Spectrum Sensing and Scheduling for {UAVs} in {UTM} Systems},
	author       = {Chintareddy, Sravan Reddy and Roach, Keenan and Cheung, Kenny and Hashemi, Morteza},
	year         = 2025,
	journal      = {IEEE Transactions on Machine Learning in Communications and Networking}
}

@article{Tan-2019-Downlink,
	title        = {Downlink {NOMA} in Multi-{UAV} Networks over Bivariate Rician Shadowed Fading Channels},
	author       = {Tan, Zheng Hui Ernest and Madhukumar, A. S. and Sirigina, Rajendra Prasad and Krishna, Anoop Kumar},
	year         = 2019,
	journal      = {Proceedings of IEEE Vehicular Technology Conference (VTC-Fall)}
}

@article{Lim-2024-Joint,
	title        = {Joint User Clustering, Beamforming, and Power Allocation for {mmWave-NOMA} With Imperfect {SIC}},
	author       = {Lim, Byungju and Yun, Won Joon and Kim, Joongheon and Ko, Young-Chai},
	year         = 2024,
	journal      = {IEEE Transactions on Wireless Communications}
}

@article{Liu-2022-Effective,
	title        = {Effective User Clustering and Power Control for Multiantenna Uplink {NOMA} Transmission},
	author       = {Liu, Ming and Zhang, Junxia and Xiong, Ke and Zhang, Mingshan and Fan, Pingyi and Letaief, Khaled B.},
	year         = 2022,
	journal      = {IEEE Trans. on Wireless Communications}
}

@article{Zeng-2019-Energy,
	title        = {Energy-Efficient Joint User-{RB} Association and Power Allocation for Uplink Hybrid {NOMA-OMA}},
	author       = {Zeng, Ming and Yadav, Animesh and Dobre, Octavia A. and Poor, H. Vincent},
	year         = 2019,
	journal      = {IEEE Internet Things Journal}
}

@article{Mohammed-2025-RIS,
	title        = {{RIS} Partitioning and User Clustering for Resilient Non-Orthogonal Multiple Access {UAV} Networks},
	author       = {Mohammed Saif and Shahrokh Valaee},
	year         = 2025,
	journal      = {arXiv:2505.00658}
}

@article{Goh-2026-Hybrid,
	title        = {Hybrid Reconfigurable Intelligent Surfaces and Unmanned Aerial Vehicles Assisted Cooperative {NOMA}},
	author       = {Goh, Chi Yen and Leow, Chee Yen and Foh, Chuan Heng and Pan, Gaofeng},
	year         = 2026,
	journal      = {IEEE Access}
}

@article{Umer-2024-Deep,
	title        = {Deep Reinforcement Learning for Trajectory and Phase Shift Optimization of Aerial {RIS} in {CoMP-NOMA} Networks},
	author       = {Umer, Muhammad and Mohsin, Muhammad Ahmed and Mahmood, Aamir and Dev, Kapal and Jung, Haejoon and Gidlund, Mikael and Hassan, Syed Ali},
	year         = 2024,
	journal      = {Proceedings of IEEE Global Communications Conference}
}

@article{Chen-2025-Secure,
	title        = {Secure Resource Allocation and Trajectory Design for {RIS} and {NOMA} Assisted Multi-{UAV} Systems},
	author       = {Chen, Jian and Zheng, Kaiqi and Jia, Jie and Deng, Yansha and Wang, Xingwei},
	year         = 2025,
	journal      = {IEEE Internet Things Journal}
}

@article{Chen-2024-CoMP,
	title        = {{CoMP} and {RIS}-Assisted Multicast Transmission in a Multi-{UAV} Communication System},
	author       = {Chen, Jian and Zhai, Kaili and Wang, Zhaolin and Liu, Yuanwei and Jia, Jie and Wang, Xingwei},
	year         = 2024,
	journal      = {IEEE Transactions on Communications}
}

@article{Huroon-2026-Energy,
	title        = {Energy-Efficient Transmission Strategy for {UAV-RIS} 2.0 Assisted Communications Using Rate Splitting Multiple Access},
	author       = {Huroon, Aamer Mohamed and Huang, Yu-Chih and Wang, Li-Chun},
	year         = 2026,
	journal      = {IEEE Trans. Wireless Commun.}
}

@article{Lin-2025-Energy,
	title        = {An Energy Effective RIS-Assisted Multi-UAV Coverage Scheme for Fairness-Aware Ground Terminals},
	author       = {Lin, Na and Wu, Tianxiong and Zhao, Liang and Hawbani, Ammar and Wan, Shaohua and Guizani, Mohsen},
	year         = 2025,
	journal      = {IEEE Trans. Green Commun. Netw.}
}

@article{Eslami-2026-Hierarchical,
	title        = {Hierarchical Resource Optimization in Multi-{UAV} {RIS}-Assisted {ISAC} Networks With Uplink {NOMA}},
	author       = {Eslami, Laleh and Kianfar, Ghazaleh and Abouei, Jamshid and Mohammadi, Arash},
	year         = 2026,
	journal      = {IEEE Internet Things Journal}
}

@article{Bhardwaj-2025-Energy,
	title        = {Energy-Efficient Node Localization in Time-Varying {UAV-RIS}-Assisted and Cluster-Based {IoT} Networks},
	author       = {Kumar Bhardwaj, Vikash and Shukla, Aagat and Pandey, Om Jee},
	year         = 2025,
	journal      = {IEEE Trans. on Network and Service Manag.}
}

@article{Saif-2025-RIS,
	title        = {{RIS} Alignment via Virtual Partitioning for Resilient Uplink Multi-{RIS}-Assisted {UAV} Communications},
	author       = {Saif, Mohammed and Valaee, Shahrokh},
	year         = 2025,
	journal      = {IEEE Transactions on Communications}
}

@article{Katwe-2022-Dynamic,
	title        = {Dynamic User Clustering and Optimal Power Allocation in {UAV}-Assisted Full-Duplex Hybrid {NOMA} System},
	author       = {Katwe, Mayur and Singh, Keshav and Sharma, Prabhat Kumar and Li, Chih-Peng and Ding, Zhiguo},
	year         = 2022,
	journal      = {IEEE Trans. Wireless Commun.}
}

@article{Katwe-2022-Energy,
	title        = {Energy Efficiency Maximization for {UAV}-Assisted Full-Duplex {NOMA} System: User Clustering and Resource Allocation},
	author       = {Katwe, Mayur and Singh, Keshav and Sharma, Prabhat Kumar and Li, Chih-Peng},
	year         = 2022,
	journal      = {IEEE Trans. Green Commun. Netw.}
}

@article{Yin-2026-Integrated,
	title        = {Integrated Deployment and Resource Allocation in Multilayer {UAV}-Enabled {NOMA} Wireless Caching Networks},
	author       = {Yin, Yue and Bouazizi, Mondher and Hu, Bintao and Gui, Guan and Ohtsuki, Tomoaki},
	year         = 2026,
	journal      = {IEEE Internet Things Journal}
}

@article{Zhong-2022-Multi,
	title        = {Multi-Agent Reinforcement Learning in {NOMA}-Aided {UAV} Networks for Cellular Offloading},
	author       = {Zhong, Ruikang and Liu, Xiao and Liu, Yuanwei and Chen, Yue},
	year         = 2022,
	journal      = {IEEE Transactions on Wireless Communications}
}

@article{Huang-2022-Resource,
	title        = {Resource Allocation for Multi-Cluster {NOMA-UAV} Networks},
	author       = {Huang, Qiulei and Wang, Wei and Lu, Weidang and Zhao, Nan and Nallanathan, Arumugam and Wang, Xianbin},
	year         = 2022,
	journal      = {IEEE Transactions on Communications}
}

@article{Na-2021-UAV,
	title        = {{UAV}-Supported Clustered {NOMA} for {6G}-Enabled Internet of Things: Trajectory Planning and Resource Allocation},
	author       = {Na, Zhenyu and Liu, Yue and Shi, Jingcheng and Liu, Chungang and Gao, Zihe},
	year         = 2021,
	journal      = {IEEE Internet Things Journal}
}

@article{Ghazikor-2026-Performance,
	title        = {Performance Analysis of {RIS}-Assisted {UAV} Communication in {NOMA} Networks},
	author       = {Ghazikor, Masoud and Nguyen, Ly V. and Hashemi, Morteza},
	year         = 2026,
	journal      = {Proc. IEEE Consumer Communications \& Networking Conf. (CCNC)}
}

@article{Kuhn-1955-Hungarian,
  author  = {Harold W. Kuhn},
  title   = {The {Hungarian} Method for the Assignment Problem},
  journal = {Naval Research Logistics Quarterly},
  year    = {1955},
}

@article{Mu-2021-Intelligent,
	title        = {Intelligent Reflecting Surface Enhanced Multi-{UAV} {NOMA} Networks},
	author       = {Mu, Xidong and Liu, Yuanwei and Guo, Li and Lin, Jiaru and Poor, H. Vincent},
	year         = 2021,
	journal      = {IEEE Journal on Selected Areas in Communications}
}

@article{Xing-2018-Optimal,
	title        = {Optimal Throughput Fairness Tradeoffs for Downlink Non-Orthogonal Multiple Access Over Fading Channels},
	author       = {Xing, Hong and Liu, Yuanawei and Nallanathan, Arumugam and Ding, Zhiguo and Poor, H. Vincent},
	year         = 2018,
	journal      = {IEEE Trans. Wireless Commun.}
}

@article{Timotheou-2015-Fairness,
	title        = {Fairness for Non-Orthogonal Multiple Access in {5G} Systems},
	author       = {Timotheou, Stelios and Krikidis, Ioannis},
	year         = 2015,
	journal      = {IEEE Signal Processing Letters}
}

@article{Tabeshnezhad-2026-Jammer,
	title        = {Jammer Mitigation in Absorptive {RIS}-Assisted Uplink {NOMA}},
	author       = {Tabeshnezhad, Azadeh and Zhu, Yuqing and Vilenskiy, Artem R. and Nguyen, Van Ly and Swindlehurst, Arnold Lee and Svensson, Tommy},
	year         = 2026,
	journal      = {IEEE Transactions on Communications}
}

\end{document}